\documentclass[12pt]{article}

\usepackage{amsmath}
\newcommand{\abs}[1]{\left\lvert#1\right\rvert}
\DeclareMathOperator{\Sim}{sim}
\DeclareMathOperator{\Dom}{dom}
\DeclareMathOperator*{\Limsup}{\varlimsup}
\DeclareMathOperator*{\Liminf}{\varliminf}
\DeclareMathOperator{\code}{code}
\usepackage{amsthm}
\usepackage{amssymb}

\usepackage{a4wide}

\newcommand{\N}{\mathbb{N}}%      \N   == \mathbb{N}
\newcommand{\Z}{\mathbb{Z}}%      \Z   == \mathbb{Z}
\newcommand{\Q}{\mathbb{Q}}%      \Q   == \mathbb{Q}
\newcommand{\R}{\mathbb{R}}%      \R   == \mathbb{R}
\newcommand{\Hm}{\mathcal{H}}%    \Hm  == \mathcal{H}
\newcommand{\Sh}{F_{\text{halt}}}
\newcommand{\Oc}{f}

\theoremstyle{plain}
\newtheorem{theorem}{Theorem}[section]

\newtheorem{corollary}[theorem]{Corollary}
\newtheorem{proposition}[theorem]{Proposition}
\theoremstyle{definition}
\newtheorem{definition}{Definition}[section]
\newtheorem{example}{Example}[section]
\theoremstyle{remark}
\newtheorem{remark}{Remark}[section]

\title{A Generalization of Chaitin's Halting\\
Probability $\Omega$ and Halting Self-Similar Sets}

\author{Kohtaro Tadaki%
\thanks{Current address:
Quantum Computation and Information Project, ERATO,
Japan Science and Technology Corporation,
Daini Hongo White Bldg.,
5-28-3, Hongo, Bunkyo-ku, Tokyo 113-0033, Japan.
E-mail: tadaki@qci.jst.go.jp}\\
\\
Department of Mathematics, Graduate School of Science,\\
Hokkaido University, Sapporo 060-0810, Japan\\
E-mail: tadaki@math.sci.hokudai.ac.jp}

\date{}

\begin{document}

\maketitle

\begin{abstract}
We generalize the concept of randomness in an infinite binary sequence
in order to characterize the degree of randomness by a real number $D>0$.
Chaitin's halting probability $\Omega$ is generalized to $\Omega^D$ whose
degree of randomness is precisely $D$.
On the basis of this generalization,
we consider the degree of randomness of each point in Euclidean space
through its base-two expansion.
It is then shown that the maximum value of such a degree of randomness
provides the Hausdorff dimension of a self-similar set that is computable
in a certain sense.
The class of such self-similar sets includes familiar fractal sets 
such as the Cantor set, von Koch curve, and Sierpi\'nski gasket.
Knowledge of the property of $\Omega^D$ allows us to show that
the self-similar subset of $[0,1]$ defined by the halting set of a
universal algorithm has a Hausdorff dimension of one.
\end{abstract}

\vspace{\baselineskip}

\textbf{Key words\/}:
algorithmic information theory, Kolmogorov complexity, randomness,
Chaitin's $\Omega$, Hausdorff dimension, self-similar set.

\vspace{\baselineskip}

\textbf{2000 Mathematics Subject Classification}\ :
Primary 68Q30; Secondary 28A80, 28A78.

\newpage

\section{Introduction}

The Kolmogorov complexity $H(s)$ of a finite binary sequence $s$ is the
size, in bits, of the shortest program for a universal algorithm $U$
to calculate $s$.
The concept of Kolmogorov complexity plays an important role in
characterizing the randomness of an infinite binary sequence.
In \cite{Chaitin 87b},
the four concepts of randomness in an infinite binary sequence
(Chaitin, weak Chaitin, Martin-L\"of, and Solovay randomness)
are considered.
These four concepts are shown to be equivalent to one another.
In this paper,
we first generalize these four concepts of randomness in order to deal
with the degree of randomness of an infinite binary sequence.
The degree of randomness is specified by a real number $D$ with
$0<D\le 1$.
As $D$ becomes larger,
the degree of randomness increases.
In the case when $D=1$,
the concept of the degree of randomness becomes the same as that of
randomness.
The relationship among the generalized concepts of randomness is
investigated.
Chaitin's halting probability $\Omega$ is an example of a random real
number.
We generalize $\Omega$ to $\Omega^D$ so that the degree of randomness
of $\Omega^D$ is precisely $D$.
Although the first $n$ bits of $\Omega$ can solve the halting problem
for a program of size not greater than $n$,
the first $n$ bits of $\Omega^D$ can solve the halting problem for a
program of size not greater than $Dn$.
Moreover,
$\Omega^D$ is infinitely differentiable as a function of $D$,
and each derivative $d^k\Omega^D/dD^k$ has the same properties as
$\Omega^D$.

On the basis of this generalization,
we next study the relationship between the degree of randomness
and Hausdorff dimension.
Hausdorff dimension is closely related to Kolmogorov complexity,
as studied by several researchers
e.g.,
\cite{Ryabko 84}, \cite{Ryabko 86}, \cite{Staiger 93},
\cite{Cai and Hartmanis 94}, and \cite{Staiger 98}.
In these previous studies however,
the normalized Kolmogorov complexity $\lim_{n \to \infty} H(x_n)/n$ of
a real number $x$ was considered,
where $x_n$ is the first $n$ bits of the base-two expansion of $x$,
and Hausdorff dimension was related to the normalized Kolmogorov
complexity.
That is to say,
in \cite{Ryabko 84}, \cite{Ryabko 86}, \cite{Staiger 93},
and \cite{Staiger 98},
the Hausdorff dimension of a subset $F$ of $\R$ was compared with
the maximum value, or supremum, over the normalized Kolmogorov
complexity $\lim_{n \to \infty} H(x_n)/n$ for all points $x$ in $F$.
(We recommend reading \cite{Staiger 93} as a monograph.)
On the other hand,
\cite{Cai and Hartmanis 94} considered the Hausdorff dimension of the
graph of the normalized Kolmogorov complexity
$\lim_{n \to \infty} H(x_n)/n$ as a function of $x$.

If an inifinite binary sequence is random,
then its normalized Kolmogorov complexity is equal to one;
however,
the converse is not necessarily true.
Thus, although the concept of normalized Kolmogorov complexity is
related to randomness it alone cannot capture randomness.
Corresponding to this fact,
the concept of the degree of randomness which we introduce in this paper
is more insightful than that of the normalized Kolmogorov complexity.
Consideration of the degree of randomness allows us to classify infinite
sequences which have the same normalized Kolmogorov complexity.
Hence,
we study the relationship between Kolmogorov complexity and Hausdorff
dimension using a more rigorous system than previous work.

We introduce six ``algorithmic dimensions'',
1st, 2nd, 3rd, 4th, upper, and lower algorithmic dimensions
as fractal dimensions for a subset $F$ of $\R^N$.
These dimensions are defined by means of Kolmogorov complexity.
On the one hand,
the 3rd, 4th, upper, and lower algorithmic dimensions are related to the
maximum value, or supremum, over the normalized Kolmogorov complexity
for all points in $F$ and were, in essence, researched
by \cite{Ryabko 86} and \cite{Staiger 93}.
On the other hand,
the 1st and 2nd algorithmic dimensions are related to the maximum value
over the degree of randomness for all points in $F$.
Therefore,
they are stronger concepts with regard to the possibilities of their
existence than the former four algorithmic dimensions.
We show that all six algorithmic dimensions are equal to the Hausdorff
dimension for any self-similar set that is computable in a certain sense.
The class of such self-similar sets includes familiar fractal sets such
as the Cantor set, von Koch curve, and Sierpi\'nski gasket.

Based on the relationship between the definition of $\Omega^D$ and the
mathematical theory of self-similar sets
(e.g., \cite{Hutchinson 81}, \cite{Hata 85}),
we define the self-similar subset $\Sh$ of $[0,1]$ by using the halting
set of a universal algorithm $U$.
We may regard $\Sh$ as the set of an endless succession of coded
messages sent through a noiseless binary communication channel.
From the property of $\Omega^D$,
it is shown that $\Sh$ has a Hausdorff dimension of one and a
zero-Lebesgue measure.

The paper is organized as follows.
In the next section,
we review some basic concepts from algorithmic information theory.
We then treat the definition of Hausdorff dimension.
Section 3 is devoted to a generalization of the concepts of randomness
in an infinite binary sequence through the introduction of a real number
$D$.
Chaitin's halting probability $\Omega$ is also generalized.
In section 4,
the six algorithmic dimensions for a subset of $\R^N$ are defined by
means of Kolmogorov complexity, and their properties are investigated.
The halting self-similar set $\Sh$ is introduced in section 5.
The Hausdorff dimension and all six algorithmic dimensions of $\Sh$ are
evaluated.

\section{Preliminary definitions}

In this section,
we first recall some basic notations from algorithmic information theory
or the theory of Kolmogorov complexity.
According to \cite{Chaitin 75},
we use some variant of Kolmogorov complexity,
i.e.,
Kolmogorov complexity based on self-delimiting programs;
we recommend reading \cite{Chaitin 75}.

$\# S$ is the cardinality of $S$ for any set $S$.
$\N \equiv \left\{0,1,2,3,\dotsc\right\}$ is the set of natural numbers,
and $\N^+$ is the set of positive integers.
$\Z$ is the set of integers, and
$\Q$ is the set of rational numbers.
$\R^N$ denotes $N$-dimensional Euclidean space,
where $\R^1 = \R$ is just the set of real numbers.
$X \equiv
\left\{
  \Lambda,0,1,00,01,10,11,000,001,010,\dotsc
\right\}$
is the set of finite binary sequences,
and $X$ is ordered as indicated.
For any $s \in X$, $\abs{s}$ is the length of $s$.
$X^\infty$ is the set of infinite binary sequences.
For any $\alpha\in X^\infty$, $\alpha_n$ is the prefix of $\alpha$ of
length $n$,
especially, $\alpha_0$ is the empty word $\Lambda$.
For any $S \subset X$,
$\mathfrak{I}(S)$ denotes the set of infinite binary sequences
beginning with a finite sequence that belongs to $S$,
i.e.,
\begin{equation}
  \mathfrak{I}(S) \equiv
  \left\{
    \>\alpha\,\in X^\infty\;\big|
    \;\exists\,n\in\N\;\,\alpha_n\in S\;
  \right\}.
\end{equation}
We write ``r.e.'' instead of ``recursively enumerable.''

A subset $S$ of $X$ is called a prefix-free set
if no sequence in $S$ is a prefix of another sequence in $S$.
For any partial recursive function $C\colon X \to X$,
the domain of $C$ is denoted by $\Dom C$,
i.e., 
$\Dom C \equiv
\left\{\>
  p\in X\bigm|C(p)\text{ is defined}
\;\right\}$.
A computer is a partial recursive function
$C\colon X \to X$ such that
$\Dom C$
is a prefix-free set.
Let $C$ be a computer. For any $s \in X$, $H_C(s)$ is defined as
\begin{equation}
  H_C(s) \equiv \min
  \left\{
    \>\abs{p}\;\big|\;p \in X\>\&\>C(p) = s\;
  \right\}
  \quad\text{(may be $\infty$)}.
\end{equation}
It is shown that there exists a computer $U$ such that for each
computer $C$ there exists a constant $\Sim(C)$ with the following
property:
if $p\in\Dom C$,
then there is a $q$ for which $U(q) = C(p)$ and
$\abs{q} \le \abs{p} + \Sim(C)$.
We choose any one of such a computer $U$ and define
$H(s) \equiv H_U(s)$,
which is referred to as the algorithmic information content of $s$,
the program-size complexity of $s$,
or the Kolmogorov complexity of $s$.
Thus $H(s)$ has the following property:
\begin{equation}
  \forall\,C:\text{computer}\quad
  H(s) \le H_C(s) + \Sim(C). \label{eq: k}
\end{equation}
We see that there is $c\in\N$ such that for any $s \neq \Lambda$,
\begin{equation}
  H(s)\le\abs{s}+2\log_2\abs{s}+c. \label{eq: fas}
\end{equation}
For any $n \in \N$, $H(n)$ is defined to be
$H($the $n$th element of $X)$.

Chaitin's halting probability $\Omega$ is defined as
\begin{equation}
  \Omega \equiv
  \sum_{p\in\Dom U}
  2^{-\abs{p}}.
\end{equation}
It is then shown that $0<\Omega < 1$.

Normally, $o(n)$ denotes any one function $f\colon \N \to \R$ such
that $\lim_{n \to \infty} f(n)/n = 0$.

Let $D$ be a real number.
$D\bmod 1$ denotes $D - \lfloor D \rfloor$,
where $\lfloor D \rfloor$ is the greatest integer less than
or equal to $D$,
and $D \bmod' 1$ denotes $D - \lceil D \rceil + 1$,
where $\lceil D \rceil$ is the smallest integer greater than
or equal to $D$.
Hence, $D\bmod 1 \in [0,1)$ but $D\bmod' 1 \in (0,1]$.
We say that $D$ is computable if the base-two expansion of $D$ can be
generated by an algorithm,
i.e.,
if there exists a total recursive function $f\colon \N^+ \to \{0,1\}$
such that
\begin{equation*}
  0.f(1) f(2) f(3) f(4) \dotsc \dotsc
\end{equation*}
is the base-two expansion of $D\bmod 1$.
The following three conditions are equivalent to one another.
\begin{enumerate}
  \item $D$ is a computable real number.
  \item If $f\colon\N^+\to\Z$ with
    $f(n)=\lfloor Dn\rfloor$ then
    $f$ is a total recursive function.
  \item There exists a total recursive function
    $f\colon\N^+\to\Z$ such that
    $\abs{D-f(n)/n} < 1/n$
    for all $n\in\N^+$.
\end{enumerate}

Let $x \in \R^N$ and use the coordinate form $x = (x^1,x^2,\dots,x^N)$.
For each $i = 1,\dots,N$ we denote $x^i\bmod 1$ in base-two notation
with infinitely many zeros:
\begin{equation}
  x^i\bmod 1 = 0.x_1^i x_2^i x_3^i \dotsc \dotsc.
\end{equation}
We then define $\code_N\colon\R^N\to X^\infty$ as
\begin{equation}
  \code_N(x) \equiv
  x_1^1 x_1^2 \dots x_1^N
  x_2^1 x_2^2 \dots x_2^N
  x_3^1 x_3^2 \dots x_3^N
  \dotsc \dotsc \dotsc.
\end{equation}
Throughout the rest of the paper,
where there is no likelihood of confusion,
$\code_N(x)$ may be denoted simply by $x$.
Thus $x_n$ is the first $n$ bits of the infinite binary sequence
$\code_N(x)$ for any $x\in\R^N$.
We will identify any point of $\R^N$ with an infinite binary sequence
in this manner.

\begin{definition}[Hausdorff dimension]
  If $U$ is any non-empty subset of $\R^N$,
  the diameter of $U$ is defined as
  $\abs{U} \equiv
  \sup
  \left\{
    \,\abs{x-y}\>\big|\>x,y\in U\,
  \right\}$.
  Suppose that $F \subset \R^N$ and $D\ge 0$.
  If $\{ U_i \}$ is a countable (or finite) collection of sets
  of diameter at most $\delta$ that cover $F$,
  i.e., $F \subset \bigcup_i U_i$ with $0<\abs{U_i}\le \delta$
  for each $i$, we say that $\{ U_i \}$ is a $\delta$-cover of $F$.
  For any $\delta >0$ we define
  \begin{equation}
    \Hm_{\delta}^D(F) \equiv
    \inf
    \biggl\{
      \;\sum_i \abs{U_i}^D\>\Big|
      \>\{ U_i \}\text{ is a $\delta$-cover of }F\;
    \biggr\}.
  \end{equation}
  We then define
  \begin{equation}
    \Hm^D(F) \equiv
    \lim_{\delta \to 0} \Hm_{\delta}^D(F).
  \end{equation}
  This limit exists for any subset $F$ of $\R^N$,
  though the limiting value can be $0$ or $\infty$.
  It is shown that $\Hm^D$ is an outer measure on $\R^N$.
  $\Hm^D$ is called $D$-dimensional Hausdorff outer measure.
  Finally,
  the Hausdorff dimension $\dim_H F$ of $F$ is defined as
  \begin{equation}
    \dim_H F \equiv \inf
    \left\{
      \,D\ge 0\,\big|\,\Hm^D(F) = 0\,
    \right\}.
  \end{equation}
\end{definition}

\bigskip
See e.g., the book \cite{Falconer 90} for a treatment of the
mathematics of Hausdorff dimension and self-similar sets.

\section{$D$-Randomness}

This section is, for all intents and purposes,
a generalization of Chapter 7 in \cite{Chaitin 87b}.

\begin{definition}[weakly Chaitin $D$-random]
  Let $D$ be a real number and $D\ge 0$,
  and let $\alpha \in X^\infty$.
  $\alpha$ is called weakly Chaitin $D$-random if
  \begin{equation}
    \exists\,c\in\R\quad\forall\,n\in\N\quad
    Dn-c \le H(\alpha_n).
  \end{equation}
\end{definition}

\bigskip
If $\mathcal{T}$ is a subset of $\N\times X$ and $i\in\N$,
we write
$\mathcal{T}_i \equiv
\left\{\,
  s\bigm|(i,s)\in\mathcal{T}
\,\right\}$.

\begin{definition}[Martin-L\"{o}f $D$-test]
  Let $\mathcal{T}\subset\N\times X$ and $D\ge 0$.
  $\mathcal{T}$ is called Martin-L\"{o}f $D$-test if
  $\mathcal{T}$ is an r.e. set and
  \begin{equation}
    \forall\,i\in\N\quad
    \sum_{s\,\in\,\mathcal{T}_i}2^{-D\abs{s}}
    \le 2^{-i}.
  \end{equation}
\end{definition}

\begin{definition}[Martin-L\"{o}f $D$-random]
  Let $D\ge 0$ and $\alpha \in X^\infty$.
  We say $\alpha$ is Martin-L\"{o}f $D$-random if
  \begin{equation}
    \forall\,\mathcal{T}:\text{Martin-L\"{o}f $D$-test}
    \quad
    \exists\,i\in\N\quad
    \alpha\notin\mathfrak{I}(\mathcal{T}_i).
  \end{equation}
\end{definition}

\bigskip
In the case where $D=1$,
the weak Chaitin $D$-randomness and Martin-L\"{o}f $D$-randomness
result in weak Chaitin randomness and Martin-L\"{o}f randomness
respectively,
which are defined in \cite{Chaitin 87b}.

\medskip
\begin{remark}
  Suppose that $D$ is a computable real number and $D\ge 0$.
  Then there exists a universal Martin-L\"{o}f $D$-test
  $\mathcal{U}^D$,
  i.e.,
  \begin{align*}
    \exists\,\mathcal{U}^D:\text{Martin-L\"{o}f $D$-test}&\quad
    \forall\,\mathcal{T}:\text{Martin-L\"{o}f $D$-test} \\
    \bigcap_{i\,\in\,\N} \mathfrak{I}(\mathcal{T}_i)&
    \subset
    \bigcap_{i\,\in\,\N} \mathfrak{I}(\mathcal{U}^D_i).
  \end{align*}
  Thus, $\alpha$ is not Martin-L\"{o}f $D$-random if and only if
  $\alpha \in
  \bigcap_{i\,\in\,\N} \mathfrak{I}(\mathcal{U}^D_i)$.
\end{remark}

\begin{theorem}\label{wcem}
  Let $D$ be a computable real number and $D\ge 0$.
  For any $\alpha\in X^\infty$,
  $\alpha$ is weakly Chaitin $D$-random
  $\Longleftrightarrow$
  $\alpha$ is Martin-L\"{o}f $D$-random.
\end{theorem}

The proof of Theorem \ref{wcem} is given in Appendix \ref{pr_wcem}.

\begin{definition}[$D$-compressible]
  Let $\alpha\in X^\infty$ and $D\ge 0$.
  We say that $\alpha$ is $D$-compressible if
  \begin{equation}
    H(\alpha_n)\le Dn+o(n),
  \end{equation}
  which is equivalent to
  \begin{equation}
    \Limsup_{n \to \infty}\frac{H(\alpha_n)}{n}\le D.
  \end{equation}
\end{definition}

\bigskip
We generalize Chaitin's halting probability $\Omega$
as follows.

\begin{definition}[Generalized halting probability]
  \begin{equation}
    \Omega^D \equiv
    \sum_{p\in\Dom U}
    2^{-\frac{\abs{p}}{D}}
    \qquad ( D > 0 ).
  \end{equation}
\end{definition}

\bigskip
Thus,
$\Omega=\Omega^1$.
If $0<D\le 1$, then $\Omega^D$ converges and $0<\Omega^D<1$,
since $\Omega^D\le \Omega<1$.

\medskip
\begin{theorem}\label{pomgd}
  Let $D$ be a real number.
  \begin{enumerate}
    \item If $0<D\le 1$ and $D$ is computable,
      then $\Omega^D$ is weakly Chaitin $D$-random and
      $D$-compressible.
    \item If $1<D$, then $\Omega^D$ diverges to infinity.
  \end{enumerate}
\end{theorem}

\begin{proof}
  (a) Suppose that $0<D\le 1$ and $D$ is
  a computable real number.

  We first show that $\Omega^D$ is weakly Chaitin $D$-random.
  The proof is a straightforward generalization of   Chaitin's
  original proof that $\Omega$ is weakly Chaitin random.
  Let $p_1,p_2,p_3,\dots$ be a recursive enumeration of the r.e. set
  $\Dom U$.
  Let $\alpha$ be the infinite binary sequence such that $0.\alpha$ is
  the base-two expansion of $\Omega^D$ with   infinitely many ones.
  Then, since $D$ is a computable real number,
  there exists a partial recursive function $\xi\colon X\to \N^+$ with
  the property that
  \begin{equation}
    0.\alpha_n<
    \sum_{i=1}^{\xi(\alpha_n)} 2^{-\frac{\abs{p_i}}{D}}.
  \end{equation}
  It is then easy to see that $Dn<\abs{p_i}$ for all $i>\xi(\alpha_n)$
  (i.e., given $\alpha_n$,
  one can calculate all programs $p$ of size not greater
  than $\lfloor Dn \rfloor$ such that $U(p)$ is defined).
  Hence, $Dn<H(s)$ for an arbitrary $s\in X$ such that $s\neq U(p_i)$
  for all $i\le \xi(\alpha_n)$.
  Therefore, given $\alpha_n$, by calculating the set
  $\left\{
      \>U(p_i)\bigm|i\le \xi(\alpha_n)\;
  \right\}$
  and picking any finite binary sequence that is not in this set,
  one can obtain an $s\in X$ such that $Dn<H(s)$.

  Thus, there exists a partial recursive function $\Psi\colon X\to X$
  with the property that
  \begin{equation}
    Dn<H(\Psi(\alpha_n)).
  \end{equation}
  Using \eqref{eq: k}, there is a natural number $c_\Psi$ such that
  \begin{equation}
    H(\Psi(\alpha_n))<H(\alpha_n)+c_\Psi.
  \end{equation}
  Therefore, $\alpha$ is weakly Chaitin $D$-random.
  It follows that $\alpha$ has infinitely many zeros,
  which implies that $\alpha=\code_1(\Omega^D)$.
  Thus, $\Omega^D$ (i.e., $\code_1(\Omega^D)$) is weakly Chaitin
  $D$-random.

  Next, we prove that $\Omega^D$ is $D$-compressible.
  We note that there exists a total recursive function
  $f\colon\N^+\times\N\to\N$ such that
  \begin{equation}
    \abs{
      \;\sum_{i=1}^k 2^{-\frac{\abs{p_i}}{D}}-2^{-n}f(k,n)\;
    }<
    2^{-n}. \label{eq: str}
  \end{equation}
  Let $\beta$ be the infinite binary sequence such that $0.\beta$ is
  the base-two expansion of the halting probability $\Omega$.

  Given $n$ and $\beta_{\lceil Dn\rceil}$
  (i.e., the first $\lceil Dn\rceil$ bits of $\beta$),
  one can find a $k_0$ with the property that
  \begin{equation}
    0.\beta_{\lceil Dn\rceil}<
    \sum_{i=1}^{k_0} 2^{-\abs{p_i}}.
  \end{equation}
  It is then easy to see that
  \begin{equation}
    \sum_{i=k_0+1}^{\infty} 2^{-\abs{p_i}}<2^{-Dn}.
  \end{equation}
  Using the inequality for real numbers 
  $a^c+b^c\le (a+b)^c\;(a,b>0,\ c\ge 1)$,
  it follows that
  \begin{equation}
    \abs{
      \;\Omega^D-\sum_{i=1}^{k_0} 2^{-\frac{\abs{p_i}}{D}}\;
    }
    <2^{-n}. \label{eq: ods}
  \end{equation}
  From \eqref{eq: str}, \eqref{eq: ods}, and
  $\abs{\Omega^D-0.\alpha_n}<2^{-n}$ it is shown that
  \begin{equation}
    \abs{\;0.\alpha_n-2^{-n}f(k_0,n)\;}<3\cdot 2^{-n}.
  \end{equation}
  Hence
  \begin{equation}
    \alpha_n=f(k_0,n),f(k_0,n)\pm 1,f(k_0,n)\pm 2,
  \end{equation}
  where $\alpha_n$ is regarded as a dyadic integer.
  Based on this, one is left with five possibilities of $\alpha_n$,
  so that one needs only $3$ bits more in order to determine $\alpha_n$.

  Thus, there exists a partial recursive function
  $\Phi\colon \N^+\times X\times X\to X$ such that
  \begin{equation}
    \forall\,n\in\N^+\quad\exists\,s\in X\quad
    \abs{s}=3\;\;\&\;\;
    \Phi(n,\beta_{\lceil Dn\rceil},s)=\alpha_n.
  \end{equation}
  From \eqref{eq: fas} it follows that
  \begin{equation}
    H(\alpha_n)
    \le |\beta_{\lceil Dn\rceil}|+o(n)
    \le Dn+o(n),
  \end{equation}
  which implies that $\Omega^D$ is $D$-compressible.

  (b) Suppose that $D>1$.
  We then choose a computable real number $d$ satisfying $D\ge d>1$.
  Let us first assume that $\Omega^d$ converges.
  Based on an argument similar to the first half of the proof of
  Theorem \ref{pomgd} (a),
  it is easy to show that $\Omega^d$ is weakly Chaitin $d$-random,
  i.e., there exists $c\in\R$ such that $dn-c\le H((\Omega^d)_n)$.
  It follows from \eqref{eq: fas} that $dn-c\le n+o(n)$.
  Dividing by $n$ and letting $n\to\infty$ we have $d\le 1$,
  which contradicts the fact $d>1$.
  Thus, $\Omega^d$ diverges to infinity.
  By noting $\Omega^d\le\Omega^D$ it is shown that $\Omega^D$
  diverges to infinity.
\end{proof}

Suppose that $0<D\le 1$ and $D$ is a computable real number.
From Theorem \ref{pomgd} (a) it follows that
$\lim_{n\to \infty} H((\Omega^D)_n)/n = D$.
Also, $(\Omega^D)_n$ solves the halting problem for a program of size
not greater than $Dn$,
as is shown in the proof of Theorem \ref{pomgd} (a).

Moreover,
as shown in the following theorem,
$\Omega^D$ is infinitely differentiable as a function of $D\in (0,1)$,
and each derived function $d^k\Omega^D/dD^k$ has the same properties
as $\Omega^D$.

\begin{theorem}\label{pomgd-dfr}
  Let $f\colon(0,1)\to\R$ with $f(D)=\Omega^D$.
  For any $p\in\Dom U$,
  let $f_p\colon(0,1)\to\R$ with $f_p(D)=2^{-\abs{p}/D}$.
  \begin{enumerate}
    \item $f$ is a function of class $C^\infty$,
      and for each $k\in\N^+$,
      \begin{equation}
        \forall\,D\in(0,1)\quad
        f^{(k)}(D) =
        \sum_{p\in\Dom U}
        f_p^{(k)}(D)
        \label{eq: termwise}
      \end{equation}
      where $f^{(k)}$ and $f_p^{(k)}$ are the $k$-th derived functions
      of $f$ and $f_p$ respectively.
    \item
      Let $k\in\N^+$ and $D$ be a computable real number in $(0,1)$. 
      Then $f^{(k)}(D)$ is weakly Chaitin $D$-random and
      $D$-compressible.
  \end{enumerate}
\end{theorem}

\begin{proof}
  It is shown that for each $k\in\N^+$,
  \begin{equation}
    f_p^{(k)}(D)=
    \frac{1}{D^k}\>Q_k\!\left(\frac{\abs{p}\ln 2}{D}\right)\,
    2^{-\frac{\abs{p}}{D}}
  \end{equation}
  where $Q_k(z)$ is the polynomial of degree $k$ with integer
  coefficients such that
  $Q_k(z)=z^k-k(k-1)z^{k-1}+\dots +(-1)^{k-1}k!z$.

  (a) We note that for each $k$ there exists $L$ such that
  if $\abs{p}\ge L$ then for any $D\in(0,1)$, $f_p^{(k)}(D)>0$.
  We wish to show by induction on $k$ that the $k$-th derived function
  $f^{(k)}$ of $f$ exists and \eqref{eq: termwise} holds.
  The result is obvious for $k=0$ from the definition of $\Omega^D$.
  Suppose that the hypothesis is true for $k=i$.
  We see that there is $L$ such that if $\abs{p}\ge L$ then
  $f_p^{(i)}(D),\ f_p^{(i+1)}(D),\ f_p^{(i+2)}(D) > 0$
  for any $D\in(0,1)$.
  Let $D\in(0,1)$. We then choose $D_0$ so that $D<D_0<1$.
  Using the mean value theorem,
  it is shown that if $\abs{p}\ge L$ then
  \begin{equation}
    f_p^{(i+1)}(D) <
    \frac{f_p^{(i)}(D_0)-f_p^{(i)}(D)}{D_0-D} <
    \frac{f_p^{(i)}(D_0)}{D_0-D}.
  \end{equation}
  Hence
  \begin{equation}
    \sum_{\abs{p}\ge L} f_p^{(i+1)}(D) \le
    \frac{1}{D_0-D} \sum_{\abs{p}\ge L} f_p^{(i)}(D_0).
  \end{equation}
  By the inductive hypothesis,
  $\sum_{p} f_p^{(i)}(D_0)$ is convergent.
  Thus $\sum_{p} f_p^{(i+1)}(D)$ is convergent for any $D\in(0,1)$.
  Since $f_p^{(i+1)}$ is a monotone increasing function for any $p$
  with $\abs{p}\ge L$,
  it is easy to see that $\sum_{p} f_p^{(i+1)}(D)$ is uniformly
  convergent on $(0,1)$ in the wider sense.
  Therefore, 
  $\sum_{p} f_p^{(i)}(D)$ is termwise differentiable,
  which implies that the hypothesis is true for $k=i+1$ as desired.

  (b) Let $k\in\N^+$.
  We then note that there exists $L\in\N$ such that if $\abs{p}>L$
  then
  \begin{equation}
    \forall\,D\in(0,1)\quad
    1 \le
    \frac{1}{D^k}\>Q_k\!\left(\frac{\abs{p}\ln 2}{D}\right)
    \label{eq: gtone}
  \end{equation}
  and
  \begin{equation}
    \left[\,
      \frac{1}{D^k}\>
      Q_k\!\left(\frac{\abs{p}\ln 2}{D}\right)
    \right]^D
  \end{equation}
  is a monotone increasing function of $D\in(0,1)$.
  Let $p_1,p_2,p_3,\dots$ be a recursive enumeration of the r.e. set
  $\left\{\,
    p\bigm|
    p\in\Dom U\;\&\>\abs{p}>L
  \right\}$.
  Also, let
  $S_L=
  \left\{\,
    p\bigm|
    p\in\Dom U\;\&\>\abs{p}\le L
  \right\}$,
  which is a finite set.
  The proof is similar to the case of $\Omega^D$.
  
  Suppose that $D$ is a computable real number in $(0,1)$.
  We then note that $\sum_{p\in S_L} f_p^{(k)}(D)$ is also a computable
  real number.

  We begin by showing that $f^{(k)}(D)$ is weakly Chaitin $D$-random.
  Let $\alpha$ be the infinite binary sequence such that $0.\alpha$ is
  the base-two expansion of $f^{(k)}(D)\bmod' 1$ with infinitely many
  ones.

  Given $\alpha_n$, one can find a $G\in\N$ with the property that
  \begin{equation}
    \lceil f^{(k)}(D) \rceil -1 + 0.\alpha_n <
    \sum_{p\in S_L} f_p^{(k)}(D) +
    \sum_{i=1}^{G} f_{p_i}^{(k)}(D).
  \end{equation}
  It is then easy to see that
  \begin{equation}
    \sum_{i=G+1}^\infty f_{p_i}^{(k)}(D) < 2^{-n}.
  \end{equation}
  Hence, from \eqref{eq: gtone}, $Dn<\abs{p_i}$ for all $i>G$.
  One can then calculate the set
  \begin{equation}
    \left\{
      \>U(p)\bigm|p\in S_L\;
    \right\}
    \cup
    \left\{
      \>U(p_i)\bigm|i\le G\;
    \right\}
  \end{equation}
  and therefore pick an $s\in X$ that is not in this set.
  It follows that $Dn<H(s)$.

  Thus, there exists a partial recursive function $\Psi\colon X\to X$
  such that
  \begin{equation}
    Dn<H(\Psi(\alpha_n)).
  \end{equation}
  Based on an argument similar to the case of $\Omega^D$,
  we see that $\alpha$ is weakly Chaitin $D$-random.
  Since $f^{(k)}(D)\bmod 1=f^{(k)}(D)\bmod' 1=0.\alpha$,
  it follows that $f^{(k)}(D)$ is weakly Chaitin $D$-random.

  Next, we prove that $f^{(k)}(D)$ is $D$-compressible.
  We note that there exists a total recursive function
  $g\colon\N^+\times\N\to\Z$ such that
  \begin{equation}
    \abs{\;
      \sum_{p\in S_L} f_p^{(k)}(D) -
      \lfloor f^{(k)}(D) \rfloor +
      \sum_{i=1}^m f_{p_i}^{(k)}(D) -
      2^{-n}g(m,n)
    \;} <
    2^{-n}. \label{eq: str-dfr}
  \end{equation}
  Let $d$ be any computable real number with $D<d<1$,
  and let $\beta$ be the infinite binary sequence such that
  $0.\beta$ is the base-two expansion of $f^{(k)}(d)\bmod 1$.
  We then note that $\sum_{p\in S_L} f_p^{(k)}(d)$ is a computable
  real number.

  Given $n$ and $\beta_{\lceil Dn/d\rceil}$
  (i.e., the first $\lceil Dn/d\rceil$ bits of $\beta$),
  one can find an $M\in\N$ with the property that
  \begin{equation}
    \lfloor f^{(k)}(d) \rfloor +
    0.\beta_{\lceil Dn/d\rceil} <
    \sum_{p\in S_L} f_p^{(k)}(d) +
    \sum_{i=1}^{M} f_{p_i}^{(k)}(d).
  \end{equation}
  It is then easy to see that
  \begin{equation}
    \sum_{i=M+1}^\infty f_{p_i}^{(k)}(d) < 2^{-Dn/d}.
  \end{equation}
  Raising both sides of this inequality to the power $d/D$ and noting
  the way of choosing $L$,
  \begin{equation}
    \begin{split}
      2^{-n} &>
      \sum_{i=M+1}^\infty
      \left[\,
        \frac{1}{d^k}\>
        Q_k\!\left(\frac{\abs{p_i}\ln 2}{d}\right)
      \right]^{d/D}
      2^{-\abs{p_i}/D} \\
      &> \sum_{i=M+1}^\infty f_{p_i}^{(k)}(D).
    \end{split}
  \end{equation}
  It follows that
  \begin{equation}
    \abs{\;
      \sum_{p\in S_L} f_p^{(k)}(D) +
      \sum_{i=1}^{M} f_{p_i}^{(k)}(D) -
      f^{(k)}(D)
    \;} <
    2^{-n}. \label{eq: ods-dfr}
  \end{equation}
  From \eqref{eq: str-dfr}, \eqref{eq: ods-dfr}, and
  \begin{equation}
    \abs{\;
      \lfloor f^{(k)}(D) \rfloor + 0.\alpha_n-f^{(k)}(D)
    \;}
    <2^{-n},
  \end{equation}
  it is shown that
  \begin{equation}
    \abs{\;\alpha_n - g(M,n)\;}
    \le 2,
  \end{equation}
  where $\alpha_n$ is regarded as a dyadic integer.

  Thus, there exists a partial recursive function
  $\Phi\colon \N^+\times X\times X\to X$ such that
  \begin{equation}
    \forall\,n\in\N^+\quad\exists\,s\in X\quad
    \abs{s}=3\;\;\&\;\;
    \Phi(n,\beta_{\lceil Dn/d\rceil},s)=\alpha_n.
  \end{equation}
  Using an argument similar to the case of $\Omega^D$,
  we see that $\alpha$ is $D/d$-compressible.
  Since $d$ is any computable real number with $D<d<1$,
  it follows that $f^{(k)}(D)$ is $D$-compressible.
\end{proof}

\begin{remark}\label{Omega}
  Suppose that $W$ is an infinite r.e. subset of $X$.
  Chaitin proved that both
  \begin{equation}
    \sum_{U(p)\in W} 2^{-\abs{p}}
  \end{equation}
  and
  \begin{equation}
    \sum_{s\in W} 2^{-H(s)}
  \end{equation}
  are weakly Chaitin $1$-random as $\Omega$.
  Corresponding to this fact,
  it is shown that both
  \begin{equation}
    \sum_{U(p)\in W} 2^{-\abs{p}/D}
    \label{eq: gp}
  \end{equation}
  and
  \begin{equation}
    \sum_{s\in W} 2^{-H(s)/D}
    \label{eq: gpc}
  \end{equation}
  have the same properties as $\Omega^D$,
  i.e.,
  the following results hold:
  (i)
  If $D>1$ then both \eqref{eq: gp} and \eqref{eq: gpc} diverge to
  infinity.
  (ii)
  As a function of $D$,
  each of \eqref{eq: gp} and \eqref{eq: gpc} is infinitely termwise
  differentiable on $(0,1)$.
  (iii)
  If $k\in\N$ and $D$ is a computable real number in $(0,1)$
  then, for each of \eqref{eq: gp} and \eqref{eq: gpc},
  the value of its $k$-th derived function at $D$ is weakly Chaitin
  $D$-random and $D$-compressible.
\end{remark}

\begin{definition}[Chaitin $D$-random]
  Let $D$ be a real number and $D\ge 0$,
  and let $\alpha \in X^\infty$.
  $\alpha$ is called Chaitin $D$-random if
  \begin{equation}
    \lim_{n \to \infty} H(\alpha_n) - Dn = \infty.
  \end{equation}
\end{definition}

\bigskip
\begin{definition}[Solovay $D$-test]
  Let $\mathcal{T} \subset \N\times X$ and $D\ge 0$.
  $\mathcal{T}$ is called Solovay $D$-test if $\mathcal{T}$ is an
  r.e. set and
  \begin{equation}
    \sum_{(i,s) \in \mathcal{T}} 2^{-D\abs{s}}
    < \infty,
  \end{equation}
  where the sum is over all $i$ and $s$ such that
  $(i,s) \in \mathcal{T}$.
\end{definition}

\begin{definition}[Solovay $D$-random]
  Let $D\ge 0$ and $\alpha \in X^\infty$.
  We say that $\alpha$ is Solovay $D$-random if
  \begin{equation}
    \forall\,\mathcal{T}:\text{Solovay $D$-test}\quad
    \exists\,m\in\N\quad
    \forall\,i>m\quad
    \alpha\notin\mathfrak{I}(\mathcal{T}_i).
  \end{equation}
\end{definition}

\bigskip
In the case where $D=1$,
the Chaitin $D$-randomness and Solovay $D$-randomness result in
Chaitin randomness and Solovay randomness respectively,
which are defined in \cite{Chaitin 87b}.

\begin{theorem}\label{ces}
  Let $D$ be a computable real number and $D\ge 0$,
  and let $\alpha\in X^\infty$.
  Then $\alpha$ is Chaitin $D$-random $\Longleftrightarrow$
  $\alpha$ is Solovay $D$-random.
\end{theorem}

The proof of Theorem \ref{ces} is immediately obtained by
generalizing the proof of Theorem R3 in \cite{Chaitin 87b}.

\begin{theorem}\label{ciwc}
  Let $D\ge 0$ and $\alpha\in X^\infty$.
  $\alpha$ is Chaitin $D$-random $\Longrightarrow$
  $\alpha$ is weakly Chaitin $D$-random.
\end{theorem}

\begin{proof}
  This is immediately apparent from the definitions.
\end{proof}

\begin{remark}\label{op_ctowc}
  The converse of Theorem \ref{ciwc} holds for $D=1$,
  because all Martin-L\"{o}f $1$-random sequences are Solovay
  $1$-random,
  as is shown in \cite{Chaitin 87b}.
  However,
  whether the converse of Theorem \ref{ciwc} also holds for any
  computable real number $D$ with $D<1$ is an open problem.
\end{remark}

\medskip
\begin{definition}[semi $D$-random]
  Let $D\ge 0$ and $\alpha \in X^\infty$.
  We say $\alpha$ is semi $D$-random if
  \begin{equation}
    D \le \Liminf_{n \to \infty} \frac{H(\alpha_n)}{n}.
  \end{equation}
\end{definition}

\bigskip
\begin{proposition}\label{wcis}
  $\alpha\text{ is weakly Chaitin $D$-random}
    \Longrightarrow
  \alpha\text{ is semi $D$-random}.$
\end{proposition}

\begin{proof}
  This is obvious from the definitions.
\end{proof}

In general, the converse of Proposition \ref{wcis} does not
necessarily hold.
For example,
although the infinite binary sequence $r_1 r_2 r_3 \dotsc \dotsc$
considered in the proof of Theorem \ref{hss} is semi $1$-random,
it is not weakly Chaitin $1$-random.

\begin{proposition}\label{eqvsdr}
  The following four conditions are equivalent to one another.
  \begin{enumerate}
    \item $\alpha$ is semi $D$-random.
    \item $\displaystyle
      Dn + o(n) \le H(\alpha_n)$.
    \item $\forall\,d\in\R\;
      \left(
        \;0\le d<D \Longrightarrow
        \alpha\text{ is Chaitin $d$-random}\;
      \right)$.
    \item $\forall\,d\in\R\;
      \left(
        \;0\le d<D \Longrightarrow
        \alpha\text{ is weakly Chaitin $d$-random}\;
      \right)$.
  \end{enumerate}
\end{proposition}

\begin{proof}
  The above equivalences follow immediately from the definitions.
\end{proof}

\section{Algorithmic Dimensions}

We introduce the six fractal dimensions which are related to the
degree of randomness or the normalized Kolmogorov complexity.

\medskip
\begin{definition}[algorithmic dimensions]
  Let $F$ be a subset of $\R^N$.
  \begin{enumerate}
  \item The 1st algorithmic dimension of $F$,
    which is denoted by $\dim_{A1} F$,
    is defined as $D\in\R$ such that
    \begin{equation}
      \forall\,x\in F\quad
      x\text{ is $\frac{D}{N}$-compressible}
    \end{equation}
    and
    \begin{equation}
      \exists\,x\in F\quad
      x\text{ is Chaitin $\frac{D}{N}$-random}.
    \end{equation}
  \item The 2nd algorithmic dimension of $F$,
    which is denoted by $\dim_{A2} F$,
    is defined as $D\in\R$ such that
    \begin{equation}
      \forall\,x\in F\quad
      x\text{ is $\frac{D}{N}$-compressible}
    \end{equation}
    and
    \begin{equation}
      \exists\,x\in F\quad
      x\text{ is weakly Chaitin $\frac{D}{N}$-random}.
    \end{equation}
  \item The 3rd algorithmic dimension of $F$,
    which is denoted by $\dim_{A3} F$,
    is defined as $D\in\R$ such that
    \begin{equation}
      \forall\,x\in F\quad
      x\text{ is $\frac{D}{N}$-compressible}
    \end{equation}
    and
    \begin{equation}
      \exists\,x\in F\quad
      x\text{ is semi $\frac{D}{N}$-random}.
    \end{equation}
  \item The 4th algorithmic dimension of $F$,
    which is denoted by $\dim_{A4} F$,
    is defined as $D\in\R$ such that
    \begin{equation}
      \forall\,x\in F\quad
      x\text{ is $\frac{D}{N}$-compressible}
    \end{equation}
    and
    \begin{equation}\label{eq: a4l}
      \forall\,d<\frac{D}{N}\quad
      \exists\,x\in F\quad
      x\text{ is Chaitin $d$-random}.
    \end{equation}
  \item The lower and upper algorithmic dimensions of $F$ are
    respectively defined as
    \begin{align}
      \underline{\dim}_A F
      &\equiv
      \sup
      \left\{
        \;D \ge 0\;\bigg|
        \;\;\exists\,x\in F
        \quad x\text{ is Chaitin $\frac{D}{N}$-random}\;\,
      \right\} \\
      &=
      \sup_{x\in F}\Liminf_{n \to \infty} \frac{H(x_n)}{n/N}
    \end{align}
    and
    \begin{align}
      \overline{\dim}_A F
      &\equiv
      \min
      \left\{
        \;D \ge 0\;\,\bigg|
        \;\;\forall\,x\in F
        \quad x\text{ is $\frac{D}{N}$-compressible}\;\,
      \right\} \\
      &=
      \sup_{x\in F}\Limsup_{n \to \infty} \frac{H(x_n)}{n/N}.
    \end{align}
  \end{enumerate}
\end{definition}

\bigskip
Although the upper and lower algorithmic dimensions always exist
unless $F$ is the empty set,
the existences of the 1st, 2nd, 3rd, and 4th algorithmic dimensions
of $F$ are nontrivial.
However,
the uniqueness of each algorithmic dimension of $F$ is trivial for
any non-empty set $F$.
Note that the condition \eqref{eq: a4l} in the definition of the
4th algorithmic dimension is equivalent to
$D \le \sup_{x\in F}\Liminf_{n \to \infty} H(x_n)/(n/N)$.
Also, from Proposition \ref{eqvsdr},
the condition ``Chaitin $d$-random''
in \eqref{eq: a4l} can be equivalently replaced
by ``weakly Chaitin $d$-random'' or ``semi $d$-random''.
Thus, we need not consider the alternative definitions which are
obtained by such replacements in the definition of the 4th algorithmic
dimension.
The `dimension' $N$ of Euclidean space $\R^N$ appears in the definition
of each algorithmic dimension.
If we identify any point in $\R^N$ with an infinite sequence over an
alphabet that consists of $2^N$ elements instead of an infinite binary
sequence,
and redefine Kolmogorov complexity using a computer whose range is the
set of finite sequences over such an alphabet,
then $N$ vanishes from these definitions.

The properties of the 3rd, 4th, upper, and lower algorithmic dimensions
were, in essence, studied by \cite{Ryabko 84} and \cite{Staiger 93}.
As more restrictive concepts,
we introduce the 1st and 2nd algorithmic dimensions which are related
to the degree of randomness instead of the normalized Kolmogorov
complexity.

\smallskip
\begin{proposition}
\label{pad}
  The algorithmic dimensions satisfy the following properties.
  \begin{enumerate}
    \item For each $k=1,2,3,4$,
      if $\dim_{Ak}F$ exists
      then $0 \le \dim_{Ak}F \le N$.
    \item \label{item:dsk}
      If $\dim_{A1}F$ exists then
      $\dim_{A2}F$ also exists and is equal to $\dim_{A1}F$.
      Similarly, for $k=2,3$,
      if $\dim_{Ak}F$ exists then
      $\dim_{A(k+1)}F$ also exists and is equal to $\dim_{Ak}F$.
    \item \label{item:nvv}
      There is $E\subset\R^N$ such that
      $\dim_{A4}E$ exists and $\dim_{A3}E$ does not exist.
      Also, there is $F\subset\R^N$ such that
      $\dim_{A3}F$ exists and $\dim_{A2}F$ does not exist.
    \item For each $k=1,2,3,4$,
      if $E\subset F$ and 
      both $\dim_{Ak}E$ and $\dim_{Ak}F$ exist
      then $\dim_{Ak}E\le\dim_{Ak}F$.
    \item For each $k=1,2,3,4$,
      if both $\dim_{Ak}E$ and $\dim_{Ak}F$ exist
      then $\dim_{Ak}(E\cup F)$ also exists and
      is equal to $\max\{\dim_{Ak}E,\dim_{Ak}F\}$.
    \item If $\dim_{A4}F_i$ exists for all $i \in \N^+$
      then
      $\dim_{A4}(\bigcup_{i=1}^\infty F_i)$ also exists and
      is equal to $\sup_{1\le i<\infty}\dim_{A4}F_i$.
    \item \label{item:open}
      If $F$ is an open subset of $\R^N$
      then $\dim_{Ak}F=N$ for $k=1,2,3,4$.
    \item If $0<D\le 1$ and $D$ is computable
      then $\dim_{Ak}\{\Omega^D\}=D$ for $k=2,3,4$.
    \item $0 \le 
      \underline{\dim}_A F \le
      \overline{\dim}_A F \le
      N$.
    \item For each $k=1,2,3,4$, if $\dim_{Ak}F$ exists
      then $\underline{\dim}_A F = \overline{\dim}_A F = \dim_{Ak}F$.
    \item If $\underline{\dim}_A F = \overline{\dim}_A F$
      then $\dim_{A4}F$ exists and these three algorithmic dimensions
      are equal to one another.
  \end{enumerate}
\end{proposition}

\begin{proof}
  These properties are obvious consequences of the definitions.
  The proof of Proposition \ref{pad} (c) is given as follows.
  Let $0<D\le 1$ and
  \begin{equation}
    E=
    \left\{\>
      \Omega^d\bigm|
      \text{$0<d<D$ and $d$ is computable}
    \;\right\}.
  \end{equation}
  Then $\dim_{A4}E=D$ but $\dim_{A3}E$ does not exist.
  Also, $\Sh$ (introduced in the next section) is an example of a set
  $F$ such that $\dim_{A3}F$ exists but $\dim_{A2}F$ does not.
  See Theorem \ref{hss}.
  Proposition \ref{pad} (g) follows from the fact that for any $s\in X$
  there is a Chaitin $1$-random infinite binary sequence whose prefix
  is $s$.
\end{proof}

Corresponding to Remark \ref{op_ctowc},
it is an open problem whether or not there is a set $F$ such that
$\dim_{A2} F$ exists and $\dim_{A1} F$ does not.

If $\dim_{A3} F$ exists, which follows from the existence of either
$\dim_{A1} F$ or $\dim_{A2} F$, 
then $\dim_{A3} F = \max_{x\in F} \lim_{n \to \infty} H(x_n)/(n/N)$,
where the maximum is over all $x\in F$ such that
$\lim_{n \to \infty} H(x_n)/(n/N)$ exists.
Note that from the definition of $\code_N$,
$x_n$ corresponds to the first $n/N$ digits of the base-two expansions
of all components of $x\in\R^N$.
This implies that $\dim_{A3} F$ is the maximum value over the
program-size complexity per digit in base-two notation for all points
in $F$.

Let $A$ be a non-empty closed subset of $\R^N$.
A transformation $S\colon A\to A$ is called a contraction on $A$ if
there is a number $c$ with $0<c<1$ such that
$\abs{S(x)-S(y)}\le c\abs{x-y}$ for all $x$, $y$ in $A$.
Let $\varphi$ denote the class of all non-empty compact subsets of $A$.
It is shown that the following theorem holds for contractions
$S_1,\dots,S_m$ on $A$ (for its proof, see e.g., \cite{Falconer 90}).

\begin{theorem}
\label{ifs}
  Let $S_1,\dots,S_m$ be contractions on $A$.
  Then there exists a unique non-empty compact set $F$ which satisfies
  \begin{equation}
    F=\bigcup_{i=1}^m S_i(F). \label{eq: is}
  \end{equation}
  Moreover, if we define a transformation $S\colon\varphi\to\varphi$ by
  \begin{equation}
    S(E)=\bigcup_{i=1}^m S_i(E)
  \end{equation}
  and write $S^k$ for the $k$-th iterate of $S$ given by
  $S^0(E)=E$, $S^k(E)=S(S^{k-1}(E))$ for $k\ge 1$,
  then
  \begin{equation}
    F=\bigcap_{k=1}^\infty S^k(E)
  \end{equation}
  for any set $E$ in $\varphi$ such that $S_i(E)\subset E$ for each $i$.
\end{theorem}

\bigskip
The unique non-empty compact set $F$ satisfying \eqref{eq: is} is
called the invariant set of the contractions $S_1,\dots,S_m$.

A contraction $S$ on $A$ is called a similarity on $A$ if there is a
number $c$ with $0<c<1$ such that $\abs{S(x)-S(y)}=c\abs{x-y}$ for all
$x$, $y$ in $A$ ($c$ is called the ratio of $S$).
The invariant set of a collection of similarities is called a
self-similar set.

Let $S_1,\dots,S_m$ be similarities on $A$.
We say that $S_1,\dots,S_m$ satisfy the open set condition if there
exists a non-empty bounded open set $V\subset A$ such that
\begin{equation}
  V \supset \bigcup_{i=1}^m S_i(V) \label{eq: osc}
\end{equation}
and $S_i(V)\cap S_j(V)=\phi\ (i\neq j)$.

\begin{theorem}
\label{main}
  Let $S_1,\dots,S_m$ be similarities on $\R^N$ with ratios
  $c_1,\dots,c_m$ respectively.
  We then note that each $S_i$ is an affine transformation,
  i.e., for each $i$ there exist $N\times N$ matrix $M_i$ and
  $v_i\in\R^N$ such that $S_i(x)=M_ix+v_i$.
  We assume that all matrix elements of $M_i$ and all components
  of $v_i$ are computable real numbers for each $i$.
  Furthermore,
  suppose that the open set condition \eqref{eq: osc} holds for
  $S_1,\dots,S_m$.
  If $F$ is the invariant set of $S_1,\dots,S_m$,
  then $\dim_{A1} F$ exists and $\dim_{A1} F = \dim_H F = D$,
  where $D$ is given by
  \begin{equation}
    \sum_{i=1}^m c_i^D = 1. \label{eq: eeq}
  \end{equation}
  Therefore, all six algorithmic dimensions of $F$ exist and are equal
  to $\dim_H F$.
\end{theorem}

\smallskip
The proof of Theorem \ref{main} is given in Appendix \ref{pr_main},
and here we only present some examples of familiar self-similar sets $F$
which are shown to satisfy $\dim_{A1} F = \dim_H F$ as a consequence of
Theorem \ref{main}.

\begin{example}
  \
  \begin{enumerate}
    \item The middle-third Cantor set is the invariant set $F$ of the
      similarities $S_1,S_2$ on $\R$ with ratios $1/3,1/3$ such that
      \begin{equation}
          S_1(x)=\frac{1}{3}x, \quad
          S_2(x)=\frac{1}{3}x+\frac{2}{3}. \label{eq: cantor}
      \end{equation}
      The open set condition \eqref{eq: osc} holds for $S_1,S_2$
      with $V$ as the open interval $(0,1)$.
      All of the real constants which appear in affine transformations
      \eqref{eq: cantor} (i.e., $0$, $1/3$, and $2/3$) are computable
      real numbers.
      Thus, by Theorem \ref{main},
      $\dim_{A1} F = \dim_H F = \log_3 2$,
      which is the solution of
      $\left(1/3\right)^D + \left(1/3\right)^D = 1$.
    \item The Sierpi\'nski gasket with vertices at the points
      $(0,0)$, $(1,0)$, and $(1/2,\sqrt{3}/2)$ is the invariant set $F$
      of the similarities $S_1,S_2,S_3$ on $\R^2$ with ratios
      $1/2,1/2,1/2$ such that
      \begin{equation}
        \begin{split}
          &S_1
          \begin{pmatrix}
            x \\
            y
          \end{pmatrix}
          =
          \begin{pmatrix}
            \frac{1}{2} &           0 \\
                      0 & \frac{1}{2}
          \end{pmatrix}
          \begin{pmatrix}
            x \\
            y
          \end{pmatrix}, \\
          &S_2
          \begin{pmatrix}
            x \\
            y
          \end{pmatrix}
          =
          \begin{pmatrix}
            \frac{1}{2} &           0 \\
                      0 & \frac{1}{2}
          \end{pmatrix}
          \begin{pmatrix}
            x \\
            y
          \end{pmatrix} +
          \begin{pmatrix}
            \frac{1}{2} \\
                     0
          \end{pmatrix}, \\
          &S_3
          \begin{pmatrix}
            x \\
            y
          \end{pmatrix}
          =
          \begin{pmatrix}
            \frac{1}{2} &           0 \\
                      0 & \frac{1}{2}
          \end{pmatrix}
          \begin{pmatrix}
            x \\
            y
          \end{pmatrix} +
          \begin{pmatrix}
                   \frac{1}{4} \\
            \frac{\sqrt{3}}{4}
          \end{pmatrix}.
        \end{split} \label{eq: ssg}
      \end{equation}
      The open set condition \eqref{eq: osc} holds for $S_1,S_2,S_3$,
      taking $V$ as the interior of the equilateral triangle with
      vertices at $(0,0)$, $(1,0)$, and $(1/2,\sqrt{3}/2)$.
      All of the real constants which appear in affine transformations
      \eqref{eq: ssg} (i.e., $0$, $1/2$, $1/4$, and $\sqrt{3}/4$) are
      computable real numbers.
      It follows from Theorem \ref{main} that
      $\dim_{A1} F = \dim_H F = \log_2 3$,
      which is the solution of
      $\left(1/2\right)^D + \left(1/2\right)^D +
      \left(1/2\right)^D = 1$.
    \item A modified von Koch curve $F\subset\R^2$ is constructed as
      follows.
      Fix a computable real number $r$ with $0<r\le 1/3$.
      Initially, consider a line segment which has endpoints $(x_1,y_1)$
      and $(x_2,y_2)$ such that all of $x_1,y_1,x_2,y_2$ are computable
      real numbers.
      Construct a curve $F$ by repeatedly replacing the middle
      proportion $r$ of each line segment by the other two sides of an
      equilateral triangle.
      (In the case where $r=1/3$, $F$ results in the von Koch curve.)
      Then we can select the four similarities $S_1,S_2,S_3,S_4$
      with ratios $\frac{1}{2}(1-r),r,r,\frac{1}{2}(1-r)$
      which have the following properties:
      (i)
      The curve $F$ is the invariant set of $S_1,\dots,S_4$.
      (ii)
      The open set condition holds for $S_1,\dots,S_4$.
      (iii)
      All of the real constants which appear in each affine
      transformation $S_i$ are computable real numbers.
      Thus, from Theorem \ref{main} we see that
      $\dim_{A1} F = \dim_H F$ = D,
      where $D$ satisfies $2r^D+2(\frac{1}{2}(1-r))^D=1$.
  \end{enumerate}
\end{example}

\section{Halting self-similar sets}

The halting self-similar set $\Sh$ is defined as
\begin{equation}
  \Sh \equiv
  \left\{\,
    0.q_1 q_2 q_3 \dotsc\bigm|
    q_i\in\Dom U\text{ for each }i
  \;\right\}.
\end{equation}
$\Sh$ is a compact subset of $[0,1]$.
Let $S_p(x)=2^{-\abs{p}}x+0.p$ for each $p\in\Dom U$.
Then $\Sh$ satisfies
\begin{equation}
  \Sh =
  \bigcup_{p\in\Dom U}
  S_p(\Sh).
\end{equation}
Thus,
since $\Dom U$ is a countably infinite set,
$\Sh$ is a self-similar set in the sense that $\Sh$ is a union of a
countably infinite number of smaller similar copies of itself.
Also, since $\Dom U$ is a prefix-free set,
the function family $\{S_p\}$ satisfies an open set condition in the
sense that there exists a non-empty bounded open set $V$
(i.e., the open interval $(0,1)$) such that
$V \supset \bigcup_p S_p(V)$ and $S_p(V)\cap S_q(V)=\phi\ (p\neq q)$.
Using the fact that $\Dom U$ is an r.e. set and not a recursive set,
it is easy to show that
$\left\{
  \>s\in X\bigm|I(s) \cap \Sh \neq \phi\;
\right\}$
is also an r.e. set and not a recursive set,
where $I(s) = [0.s, 0.s + 2^{-\abs{s}})$.

\medskip
\begin{remark}
  As considered in \cite{Chaitin 75},
  think of $U$ as decoding equipment at the receiving end of
  a noiseless binary communication channel.
  Regard its programs (i.e., finite binary sequences in $\Dom U$)
  as code words and regard the result of the computation by $U$ as the
  decoded message.
  Since $\Dom U$ is a prefix-free set,
  such code words form what is called an ``instantaneous code,''
  so that successive messages sent through the channel can be separated.
  Then $\Sh$ is the set of $x\in [0,1]$ such that the base-two expansion
  of $x$ is an endless succession of coded messages sent through the
  channel.
\end{remark}

\begin{theorem}\label{hss}
  $\dim_H \Sh = 1$ and $\mathcal{L}^1(\Sh) = 0$,
  where $\mathcal{L}^1$ is Lebesgue measure on $\R$.
  Neither $\dim_{A1} \Sh$ nor $\dim_{A2} \Sh$ exists,
  but $\dim_{A3} \Sh = \dim_{A4} \Sh = 1$.
\end{theorem}

\begin{proof}
  To begin with, we show that $\dim_H\Sh = 1$.
  Let $p_1,p_2,p_3,\dots$ be a recursive enumeration of
  the r.e. set $\Dom U$,
  and let
  \begin{equation}
    P_m =
    \left\{\,
      0.q_1 q_2 q_3 \dotsc\bigm|
      q_i\in\{p_1,p_2,\dots,p_m\}\text{ for each }i
    \;\right\}.
  \end{equation}
  Then $P_m$ is the invariant set of $S_{p_1},S_{p_2},\dots,S_{p_m}$.
  Since the open set condition \eqref{eq: osc} holds for
  $S_{p_1},S_{p_2},\dots,S_{p_m}$,
  from Theorem \ref{sss} in Appendix \ref{pr_main} it is shown that
  $\dim_H P_m = D_m$,
  where $D_m$ is given by
  \begin{equation}
    \sum_{i=1}^m 2^{-D_m\abs{p_i}} = 1.
  \end{equation}

  Now, from the definition of $\Omega^D$,
  \begin{equation}
    \Omega^\frac{1}{D}=\sum_{i=1}^\infty 2^{-D\abs{p_i}}.
    \label{eq: invomgd}
  \end{equation}
  From Theorem \ref{pomgd} (b),
  this sum diverges to infinity for each $D\in(0,1)$.
  Hence, given $\varepsilon>0$, for all sufficiently large $m$
  \begin{equation}
    \sum_{i=1}^m 2^{-(1-\varepsilon)\abs{p_i}}>1
  \end{equation}
  and
  \begin{equation}
    \sum_{i=1}^m 2^{-1\cdot\abs{p_i}}<\Omega^1<1,
  \end{equation}
  which implies that $1-\varepsilon<D_m<1$.
  Thus,
  $\lim_{m\to\infty} D_m = 1$.
  Since $P_m\subset\Sh$, it follows that $\dim_H\Sh = 1$.

  Second, we prove that $\mathcal{L}^1(\Sh) = 0$.
  We see that for each $n\in\N^+$,
  \begin{equation}
    \begin{split}
      \mathcal{L}^1(\Sh)
      &\le
      \mathcal{L}^1
      \left(
        \left\{\>
          0.q_1 \dots q_n \alpha \bigm|
          q_1, \dots, q_n \in \Dom U\ \&\ 
          \alpha \in X^\infty
        \,\right\}
      \right) \\
      &=
      \sum_{q_1,\dots,q_n\in\Dom U}
      2^{-\abs{q_1 \dots q_n}} \\
      &=
      (\Omega^1)^n.
    \end{split}
  \end{equation}
  Since $0<\Omega^1<1$,
  letting $n\to\infty$ gives $\mathcal{L}^1(\Sh) = 0$.

  Third, we prove that $\dim_{A3}\Sh = \dim_{A4}\Sh = 1$.
  Fix a weakly Chaitin $1$-random sequence $\beta$ such as the
  base-two expansion of $\Omega$.
  For each $k\in\N^+$, let $r_k$ be any one of the shortest $q$ such
  that $U(q)$ is equal to the $k$ bits sequence from $(k-1)k/2+1$th bit
  to $k(k+1)/2\,$th bit of $\beta$.
  Also, let $y = 0.r_1 r_2 r_3 \dotsc \dotsc$.
  It follows that $y \in \Sh$.

  Given $y_n$,
  one can find $r_1,r_2,r_3,\dots,r_m, t$ such that
  $y_n = r_1 r_2 r_3 \dots r_m t$ and $t$ is a proper prefix of
  $r_{m+1}$,
  possibly $t=\Lambda$.
  One can then calculate $\beta_{m(m+1)/2}$ from
  $r_1, r_2, r_3, \dots, r_m$.
  Hence, there exists a partial recursive function
  $\Psi\colon X\to X$ such that for each $n\in\N^+$,
  $\Psi(y_n) = \beta_{m(m+1)/2}$ where $m$ is the greatest integer with
  the property that $\abs{r_1 r_2 r_3 \dots r_m}\le n$.
  Using \eqref{eq: k},
  it is easy to show that there is $d\in\N$ such that
  \begin{equation}
    \frac{m(m+1)}{2}-d \le H(y_n). \label{eq: sita}
  \end{equation}

  However, using \eqref{eq: fas},
  $\abs{r_k}\le k+2\log_2 k+c$ for any $k\in\N^+$.
  Thus
  \begin{equation}
    \begin{split}
      n &< \abs{r_1 r_2 r_3 \dots r_m r_{m+1}} \\
      &\le \frac{(m+1)(m+2)}{2}+\log_2 (m+1)!+c(m+1).
    \end{split}
  \end{equation}
  Since letting $n\to\infty$ implies $m\to\infty$,
  it follows that
  \begin{equation}
    1 \le \Liminf_{n\to\infty} \frac{m(m+1)}{2n}.
  \end{equation}
  Combining with \eqref{eq: sita} this implies that $y$ is semi
  $1$-random.

  Now, it follows from \eqref{eq: fas} that $x$ is $1$-compressible
  for all $x\in\R$.
  Thus, $\dim_{A3}\Sh = 1$, which shows,
  from Proposition \ref{pad} (b), that $\dim_{A4}\Sh = 1$.

  Finally, we show that neither $\dim_{A1}\Sh$ nor $\dim_{A2}\Sh$
  exists.
  If $\dim_{A2}\Sh$ exists then, by Proposition \ref{pad}
  (b),
  $\dim_{A2}\Sh = \dim_{A3}\Sh = 1$,
  which implies that there is $x\in\Sh$ such that $x$ is Martin-L\"{o}f
  $1$-random.
  However,
  we will show that $x$ is not Martin-L\"{o}f $1$-random for any $x\in\Sh$.

  Choosing $a\in\Q$ with $\Omega^1 < a < 1$ it follows that
  \begin{equation}
    \sum_{q_1, \dots, q_n \in \Dom U} 2^{-\abs{q_1 \dots q_n}}
    < a^n.
  \end{equation}
  Let $f\colon\N\to\N$ be a total recursive function such that
  $a^{f(i)} \le 2^{-i}$ for all $i\in\N$.
  The r.e. set
  \begin{equation}
    \mathcal{T} \equiv
    \left\{\,
      (i, q_1 q_2 \dots q_{f(i)}) \bigm|
      i \in \N
      \ \&\
      q_1, q_2, \dots, q_{f(i)} \in \Dom U
    \,\right\}
  \end{equation}
  is then Martin-L\"{o}f $1$-test.
  For any $x\in\Sh$,
  $\forall\,i\in\N\;\,x\in\mathfrak{I}(\mathcal{T}_i)$ and hence $x$
  is not Martin-L\"{o}f $1$-random.

  Thus, $\dim_{A2}\Sh$ does not exist.
  From Proposition \ref{pad} (b),
  $\dim_{A1}\Sh$ also does not exist.
\end{proof}

\medskip
We say that $\Oc:X\to X$ is an optimal code if for each $s\in X$,
$\Oc(s)$ is one of the shortest program for $U$ to calculate $s$,
i.e., $U(\Oc(s))=s$ and $\abs{\Oc(s)}=H(s)$.
For any optimal code $\Oc$ and $W \subset X$,
$F_{\text{opt}}(\Oc,W)$ is defined as
\begin{equation}
  F_{\text{opt}}(\Oc,W)
  \equiv
  \left\{\,
    0.\Oc(s_1) \Oc(s_2) \Oc(s_3) \dotsc \dotsc \bigm|
    s_i\in W\text{ for each }i
  \;\right\}.
\end{equation}
The following theorem, which is similar to Theorem \ref{hss}, then holds.

\begin{theorem}\label{fopt}
  Suppose that $\Oc$ is an optimal code and
  $W$ is an infinite r.e. subset of $X$.
  Then
  \begin{equation*}
    \dim_{A3} F_{\text{opt}}(\Oc,W) =
    \dim_{A4} F_{\text{opt}}(\Oc,W) =
    \dim_H F_{\text{opt}}(\Oc,W) =1
  \end{equation*}
  and $\mathcal{L}^1(F_{\text{opt}}(\Oc,W)) = 0$.
  However,
  neither $\dim_{A1} F_{\text{opt}}(\Oc,W)$ nor
  $\dim_{A2} F_{\text{opt}}(\Oc,W)$ exists.
\end{theorem}

\begin{proof}
  The sum
  \begin{equation}
    \sum_{s\in W} 2^{-H(s)/D} \label{eq: opt_sum}
  \end{equation}
  diverges to infinity for any $D>1$,
  as we mentioned in Remark \ref{Omega}.
  Thus, using an argument similar to the case of $\Sh$,
  we see that $\dim_H F_{\text{opt}}(\Oc,W) = 1$.

  Note that $F_{\text{opt}}(\Oc,W) \subset \Sh$.
  Thus,
  $\mathcal{L}^1(F_{\text{opt}}(\Oc,W)) \le \mathcal{L}^1(\Sh) = 0$.

  Next, we prove $\dim_{A3} F_{\text{opt}}(\Oc,W) = 1$.
  Fix a weakly Chaitin $1$-random sequence $\beta$.
  For each $k\in\N^+$, let $r_k$ be any one of the shortest
  $q$ such that $U(q)$ is equal to the $k$ bits sequence
  from $(k-1)k/2+1$th bit to $k(k+1)/2\,$th bit of $\beta$.
  Since $W$ is an infinite r.e. set,
  there exists a one-to-one total recursive function
  $\xi\colon X\to W$.
  Let $y = 0.\Oc(\xi(r_1))$
  $\Oc(\xi(r_2))\,\Oc(\xi(r_3)) \dotsc \dotsc$. %%%%% modeq
  It is then shown that $y \in F_{\text{opt}}(\Oc,W)$ and there is
  $c_\xi \in \N$ such that for any $k\in\N^+$,
  $\abs{\Oc(\xi(r_k))} \le k+2\log_2 k+c_\xi$.
  Moreover, one can calculate $\beta_{m(m+1)/2}$ from
  $\Oc(\xi(r_1)),\Oc(\xi(r_2)),\Oc(\xi(r_3)),\dots,\Oc(\xi(r_m))$.
  Thus, making an argument similar to the case of $\Sh$ it is shown that
  $y$ is semi $1$-random.
  Hence,
  $\dim_{A3} F_{\text{opt}}(\Oc,W) = 1$,
  and therefore $\dim_{A4} F_{\text{opt}}(\Oc,W) = 1$.

  As was shown in the proof of Theorem \ref{hss},
  there is no Martin-L\"{o}f $1$-random sequence in $\Sh$.
  Since $F_{\text{opt}}(\Oc,W) \subset \Sh$,
  there is no Martin-L\"{o}f $1$-random sequence in
  $F_{\text{opt}}(\Oc,W)$.
  Thus, neither $\dim_{A1} F_{\text{opt}}(\Oc,W)$ nor
  $\dim_{A2} F_{\text{opt}}(\Oc,W)$ exists.
\end{proof}

For each $W \subset X$,
we define
\begin{equation}
  F_{\text{halt}}(W)
  \equiv
  \left\{\,
    0.q_1 q_2 q_3 \dotsc \dotsc \bigm|
    U(q_i)\in W\text{ for each }i
  \;\right\},
\end{equation}
which is a generalization of $\Sh$,
i.e., $\Sh = F_{\text{halt}}(X)$.
Note that
\begin{equation}\label{eq: ohh}
  F_{\text{opt}}(\Oc,W)\subsetneqq F_{\text{halt}}(W) \subset \Sh
\end{equation}
for any optimal code $\Oc$ and any infinite r.e. set $W \subset X$.
The following generalization of Theorem \ref{hss} holds.

\begin{theorem}\label{fhalt}
  Suppose that $W$ is an infinite r.e. subset of $X$.
  Then
  \begin{equation*}
    \dim_{A3} F_{\text{halt}}(W) =
    \dim_{A4} F_{\text{halt}}(W) =
    \dim_H F_{\text{halt}}(W) =1
  \end{equation*}
  and $\mathcal{L}^1(F_{\text{halt}}(W)) = 0$.
  However,
  neither $\dim_{A1} F_{\text{halt}}(W)$ nor
  $\dim_{A2} F_{\text{halt}}(W)$ exists.
\end{theorem}

\begin{proof}
  This follows immediately from Theorem \ref{hss}, Theorem \ref{fopt},
  and \eqref{eq: ohh}.
\end{proof}

\smallskip
\begin{remark}
  \cite{Ryabko 84} showed that for any r.e. set $L\subset X$,
  \begin{equation}
    \underline{\dim}_A L^{\,\infty}
    = \dim_H L^{\,\infty}
    \label{eq: ryabko_lgg}
  \end{equation}
  where
  $L^{\,\infty} =
  \left\{\,
    0.l_1 l_2 l_3 \dotsc\bigm|
    l_i\in L\text{ for each }i
  \;\right\}$
  (see also \cite{Staiger 93}).
  Suppose that $\Oc$ is an optimal code and
  $W$ is an infinite r.e. subset of $X$.
  Since
  $\left\{
    \>q\in X\bigm|U(q)\in W\;
  \right\}$
  is an r.e. set,
  using \eqref{eq: ryabko_lgg} and $\dim_{A3} F_{\text{halt}}(W) = 1$
  we immediately see that $\dim_H F_{\text{halt}}(W) = 1$.
  On the other hand,
  since
  $\left\{
    \>\Oc(s)\bigm|s\in W\;
  \right\}$
  is not an r.e. set,
  it would seem difficult to prove $\dim_H F_{\text{opt}}(\Oc,W) =1$
  directly from \eqref{eq: ryabko_lgg} and
  $\dim_{A3} F_{\text{opt}}(\Oc,W) = 1$.
  In Theorem \ref{fopt},
  using the property of the sum \eqref{eq: gpc},
  we proved $\dim_H F_{\text{opt}}(\Oc,W) =1$.
\end{remark}

\appendix
\section{Appendix}

\subsection{The proof of Theorem \ref{wcem}}
\label{pr_wcem}

In the case where $D=1$, Theorem \ref{wcem} results in Theorem R1 in
\cite{Chaitin 87b}.
The proof of Theorem \ref{wcem} is a straightforward generalization of
the proof of Theorem R1 given in \cite{Chaitin 87b}.
We need the following two theorems shown in \cite{Chaitin 75}.

\begin{theorem}\label{kie}
  Let both $f\colon\N\to X$ and $g\colon\N\to\N$ be total recursive
  functions.
  Suppose that
  \begin{equation}
    \sum_{n=0}^\infty 2^{-g(n)} \le 1.
  \end{equation}
  Then there exists a computer $C$ such that
  \begin{equation}
    H_C(s)=\min_{f(n)=s} g(n).
  \end{equation}
\end{theorem}

\begin{theorem}\label{mcfs}
There is $c\in \N$ such that for any $n\in\N$ and $k\in\Z$,
  \begin{equation}
    \#
    \left\{
      \,s\in X\bigm|\abs{s}=n\;\&\;H(s)<k\,
    \right\} <
    2^{k-H(n)+c}.
  \end{equation}
\end{theorem}

\bigskip
Theorem \ref{kie} and Theorem \ref{mcfs} are Theorem 3.2 and
Theorem 4.2 (b) in \cite{Chaitin 75}, respectively.

The proof of Theorem \ref{wcem} is as follows.

\begin{proof}[Proof of Theorem \ref{wcem}]
  Suppose that $D$ is a computable real number and $D\ge 0$.
  Let $f\colon\N\to\N$ with $f(n)=\lfloor Dn\rfloor$.
  Then $f$ is a total recursive function.\\
  ($\neg$ (\textit{weak Chaitin\/})
  $\Longrightarrow$
  $\neg$ \textit{Martin-L\"{o}f\/})

  $\neg$ (weak Chaitin) says that
  for any $k\in\N$ there is $n\in\N$ such that
  $H(\alpha_n)<f(\abs{\alpha_n})-k$.
  Let
  $\mathcal{T} =
  \left\{
    \,(k,s)\in\N\times X
    \bigm|H(s)<f(\abs{s})-k-c\,
  \right\}$
  for the natural number $c$ which is referred to in Theorem \ref{mcfs}.
  Then $\alpha\in\mathfrak{I}(\mathcal{T}_k)$ for any $k\in\N$.

  However,
  it follows from Theorem \ref{mcfs} that
  $\#\left\{
    \,s\in\mathcal{T}_k\bigm|\abs{s}=n\,
  \right\} \le
  2^{Dn-H(n)-k}$
  for any $k,n\in\N$.
  Hence, for any $k\in\N$ we get
  \begin{equation}
    \begin{split}
      \sum_{s \in \mathcal{T}_k} 2^{-D\abs{s}}
      &= \sum_{n=0}^\infty
      \#\left\{
        \,s\in\mathcal{T}_k\bigm|\abs{s}=n\,
      \right\}
      2^{-Dn} \\
      &\le 2^{-k} \left(\sum_{n=0}^\infty 2^{-H(n)}\right)
      \le 2^{-k} \Omega
      < 2^{-k}.
    \end{split}
  \end{equation}
  Since $f$ is a total recursive function,
  $\mathcal{T}$ is an r.e. set.
  Thus, $\mathcal{T}$ is Martin-L\"{o}f $D$-test,
  and hence $\alpha$ is not Martin-L\"{o}f $D$-random.\\
  ($\neg$ \textit{Martin-L\"{o}f\/}
  $\Longrightarrow$
  $\neg$ (\textit{weak Chaitin\/}))

  Suppose that there exists a Martin-L\"{o}f $D$-test $\mathcal{T}$
  such that $\alpha\in\mathfrak{I}(\mathcal{T}_n)$ for any $n\in\N$.
  Then
  \begin{equation}
    \begin{split}
      \sum_{n=2}^\infty\sum_{s\in\mathcal{T}_{n^2}}
      2^{-[f(\abs{s})-n]} &\le
      \sum_{n=2}^\infty
      \Bigl(
        2^{n+1}\sum_{s\in\mathcal{T}_{n^2}}
        2^{-D\abs{s}}
      \Bigr) \\
      &\le \sum_{n=2}^\infty 2^{-n^2+n+1} \le 1.
    \end{split}
  \end{equation}
  Since $\mathcal{T}$ is an r.e. set,
  there exists a bijective total recursive function $g$ from $\N$ to
  the set
  $\left\{\,
    (n,s)\bigm|
    n\ge 2\;\;\&\;\,s\in\mathcal{T}_{n^2}
  \,\right\}$.
  Let $n(k)$ and $s(k)$ be total recursive functions such that
  $g(k)=(n(k),s(k))$ for all $k\in\N$.
  Then
  \begin{equation}
    \sum_{k=0}^\infty 2^{-[f(\abs{s(k)})-n(k)]} =
    \sum_{n=2}^\infty\sum_{s\in\mathcal{T}_{n^2}}
    2^{-[f(\abs{s})-n]} \le 1.
  \end{equation}
  Since $f$ is a total recursive function,
  by Theorem \ref{kie}, there is a computer $C$ such that
  \begin{equation}
    H_C(s)=\min_{s(k)=s} \left\{f(\abs{s(k)})-n(k)\right\}.
  \end{equation}
  Using \eqref{eq: k}, it follows that
  \begin{equation}
    n\ge 2\;\;\&\;\,s\in\mathcal{T}_{n^2}\quad
    \Longrightarrow\quad
    H(s)\le D\abs{s}-n+\Sim(C).
  \end{equation}
  Thus, since $\alpha\in\mathfrak{I}(\mathcal{T}_{n^2})$
  for all $n\ge 2$,
  we see that for all $n\ge 2$ there exists $k\in\N$ such that
  \begin{equation}
    H(\alpha_k) \le
    D\abs{\alpha_k}-n+\Sim(C) =
    Dk-n+\Sim(C),
  \end{equation}
  which implies that $\alpha$ is not weakly Chaitin $D$-random.
\end{proof}

\subsection{The proof of Theorem \ref{main}}
\label{pr_main}

\medskip
For each $D\ge 0$,
we define $T^1_D$ and $T^2_D$ by
\begin{align}
  &T^1_D \equiv
  \left\{\,
    x\in\R^N\,\bigg|
    \,x\text{ is not Chaitin $\frac{D}{N}$-random}
  \,\right\} \\
  &T^2_D \equiv
  \left\{\,
    x\in\R^N\,\bigg|
    \,x\text{ is not weakly Chaitin $\frac{D}{N}$-random}
  \,\right\}.
\end{align}

\begin{theorem}
\label{lbt}
  Let $D\ge 0$.
  If $D$ is a computable real number,
  then $T^1_D$ and $T^2_D$ are Borel sets and
  $\Hm^D(T^1_D)=\Hm^D(T^2_D)=0$.
\end{theorem}

\begin{proof}
  In the case that $D=0$, the results are obvious from the fact that
  $T^1_0=T^2_0=\phi$.
  Thus we assume that $D>0$.

  We first show that $\Hm^D(T^1_D)=0$.
  Suppose that $\mathcal{T}$ is Solovay $D/N$-test.
  For any $s\in X$, we write
  $U(s)=
  \left\{
    \>x\in[0,1)^N\bigm|
    s\text{ is the prefix of }x\;
  \right\}$.
  It follows that
  $\sum_{(i,s) \in \mathcal{T}} \abs{U(s)}^D
  < \infty$.
  Also, we let
  $\Gamma(\mathcal{T}) =
  \left\{
    \>x\in[0,1)^N\bigm|
    \forall\,m\in\N\;\>\exists\,i>m\;\>
    x\in\mathfrak{I}(\mathcal{T}_i)\;
  \right\}$.
  Then,
  for any $\delta,\, \varepsilon > 0$,
  it is shown that there exists a $\delta$-cover $\{U(s_k)\}_k$ of
  $\Gamma(\mathcal{T})$ such that $\sum_k \abs{U(s_k)}^D < \varepsilon$.
  Hence,
  $\Hm^D_\delta(\Gamma(\mathcal{T}))<\varepsilon$.
  It follows that $\Hm^D(\Gamma(\mathcal{T}))=0$.

  On the other hand,
  Theorem \ref{ces} implies that
  $T^1_D\cap[0,1)^N=\bigcup\Gamma(\mathcal{T})$,
  where the union is over all Solovay $D/N$-test $\mathcal{T}$.
  Since there are only countably many Solovay $D/N$-tests,
  it follows that $\Hm^D(T^1_D\cap[0,1)^N)=0$.
  By noting the fact that Hausdorff outer measures are translation
  invariant
  (i.e., $\Hm^D(F+z)=\Hm^D(F)$, where
  $F+z =
  \left\{
    x+z\bigm|x\in F
  \right\}$
  ),
  we see that $\Hm^D(T^1_D)=0$.

  From Theorem \ref{ces} it follows that
  \begin{equation}
    T^1_D =
    \bigcup_\mathcal{T}
    \bigcap_{n=0}^\infty
    \bigcup_{i=n}^\infty
    \left\{
      \>x\in\R^N\bigm|
      \code_N(x)\in\mathfrak{I}(\mathcal{T}_i)\;
    \right\},
  \end{equation}
  where the leftmost union is over all Solovay $D/N$-test $\mathcal{T}$.
  Thus, $T^1_D$ is a Borel set.

  Similarly, using Theorem \ref{wcem},
  it is shown that $\Hm^D(T^2_D)=0$ and $T^2_D$ is a Borel set.
\end{proof}

\medskip
In the case where $N=1$ and $D=1$,
$\Hm^D(T^1_D)=\Hm^D(T^2_D)=0$ states the well-known fact that the set
of all non-random real numbers has a zero-Lebesgue measure.

For each $D\ge 0$,
we define $T^3_D$ by
\begin{equation}
  T^3_D \equiv
  \left\{
    \,x\in\R^N\,\bigg|
    \,x\text{ is not semi $\frac{D}{N}$-random}\,
  \right\}.
\end{equation}

\begin{corollary}[Staiger \cite{Staiger 93}, Cai and Hartmanis
\cite{Cai and Hartmanis 94}]\label{crlbt}
  If $D \ge 0$ then $T^3_D$ is a Borel set and $\Hm^D(T^3_D)=0$.
\end{corollary}

\begin{proof}
  In the case that $D=0$, the results are obvious from the fact that
  $T^3_0=\phi$.
  Thus we assume that $D>0$.
  Let $D_1, D_2, \dotsc$ be a sequence of computable real numbers such
  that $\lim_{n \to \infty} D_n=D$ and for any $n$, $D_n < D$.
  Using the equivalency between (a) and (c) in Proposition \ref{eqvsdr},
  we see that $T^3_D=\bigcup_{n=1}^\infty T^1_{D_n}$.
  Hence, by Theorem \ref{lbt}, $T^3_D$ is a Borel set.
  Since $\Hm^D$ is non-increasing with $D$,
  it follows that $\Hm^D(T^3_D)=0$.
\end{proof}

\begin{corollary}[Ryabko \cite{Ryabko 84}]\label{ryabko}
  $\dim_H F \le \underline{\dim}_A F$,
  and for each $k=1,2,3,4$, if $\dim_{Ak}F$ exists
  then $\dim_H F \le \dim_{Ak}F$.
\end{corollary}

\begin{proof}
  Let $D=\dim_H F$.
  Since the results are trivial for $D=0$,
  we assume that $D>0$.
  For any $\varepsilon>0$,
  we choose a computable real number $d$ such that
  $D-\varepsilon\le d<D$.
  From Theorem \ref{lbt} it follows that
  $\Hm^d(F\backslash T^1_d) = \Hm^d(F) > 0$.
  Hence, $F\backslash T^1_d\neq\phi$ and therefore there is $x\in F$
  such that $x$ is Chaitin $d/N$-random.
  Thus, we see that $D-\varepsilon \le \underline{\dim}_A F$
  for any $\varepsilon > 0$,
  from which the results are easily produced.
\end{proof}

\begin{definition}[r.e. condition]
  Suppose that $F$ is a subset of $\R^N$.
  We first define
  \begin{equation}
    \begin{split}
      F\bmod 1
      \equiv
      \bigl\{
        (x^1\bmod 1, x^2\bmod 1, \dots&, x^N\bmod 1)\bigm| \\
        (&x^1, x^2,  \dots, x^N) \in F\;
      \bigr\}.
    \end{split}
  \end{equation}
  For any $s\in X$,
  we define $I(s) \equiv [0.s, 0.s + 2^{-\abs{s}})$ and
  $\hat{I}(s) \equiv
  [0.s-2^{-\abs{s}}, 0.s + 2^{-\abs{s}} + 2^{-\abs{s}})\bmod 1$.
  We also generalize $I(s)$ and $\hat{I}(s)$ to intervals on $\R^N$
  by the following manner.
  \begin{align}
    &I(s_1, s_2, \dots, s_N) \equiv
    I(s_1) \times I(s_2) \times \dots \times I(s_N), \\
    &\hat{I}(s_1, s_2, \dots, s_N) \equiv
    \hat{I}(s_1) \times \hat{I}(s_2) \times \dots
    \times \hat{I}(s_N).
  \end{align}
  Finally, we define
  \begin{equation}
    \begin{split}
      \mathfrak{M}(F)
      \equiv
      \bigl\{
        (s_1, \dots, s_N)\bigm|
        \abs{s_1} =& \dots = \abs{s_N}\;
        \& \\
        \;I(s_1&, \dots, s_N) \cap (F\bmod 1) \neq \phi\;
      \bigr\}
    \end{split}
  \end{equation}
  and
  \begin{equation}
    \begin{split}
      \hat{\mathfrak{M}}(F)
      \equiv
      \bigl\{
        (s_1, \dots, s_N)\bigm|
        \abs{s_1} =& \dots = \abs{s_N}\;
        \& \\
        \;\hat{I}(s_1&, \dots, s_N) \cap (F\bmod 1) \neq \phi\;
      \bigr\}.
    \end{split}
  \end{equation}

  We say that $F$ satisfies the r.e. condition if
  there exists an r.e. set $L$ such that
  $\mathfrak{M}(F) \subset L \subset \hat{\mathfrak{M}}(F)$.
\end{definition}

\medskip
The meaning of the r.e. condition is as follows.
First we note that $F\bmod 1 \subset [0,1)^N$.
For all $n\in\N$, divide $[0,1)^N$ into $2^{Nn}$ pieces of $N$
dimensional subintervals in the form $I(s_1, s_2, \dots, s_N)$
with $\abs{s_1} = \abs{s_2} = \dots = \abs{s_N} = n$.
Then, intuitively,
the r.e. condition is that all of the subintervals $I$'s intersecting
$F\bmod 1$ and some of the subintervals neighboring to these $I$'s form
an r.e. set.

Let $F$ be a bounded subset of $\R^N$, and let $N_\delta(F)$ be the
smallest number of closed balls of radius $\delta$ that cover $F$.
The upper box-counting dimension of $F$ is defined as
\begin{equation}
  \overline{\dim}_B F \equiv
  \Limsup_{\delta \to 0} \frac{\log N_\delta(F)}{-\log\delta}.
\end{equation}

\begin{theorem}[Kolmogorov]\label{ubt}
  If $F$ is a bounded subset of $\R^N$
  and satisfies the r.e. condition,
  then $x$ is $(\overline{\dim}_B F)/N$-compressible for any $x\in F$,
  i.e.,
  \begin{equation}
    \forall\,x\in F\quad\forall\,n\in\N\quad
    H(x_n)\le \frac{\overline{\dim}_B F}{N}n + o(n).
  \end{equation}
\end{theorem}

\begin{proof}
  The essential part of the proof is due to Kolmogorov.

  For each $m\in\N$,
  we consider the collection of cubes in
  the $2^{-m}$-coordinate mesh of $\R^N$,
  i.e., the collection of sets of the form
  \begin{equation}
    [l_1 2^{-m}, (l_1+1) 2^{-m}) \times \dots \times
    [l_N 2^{-m}, (l_N+1) 2^{-m}),
  \end{equation}
  where $l_1, \dots, l_N$ are integers.
  Let $M_m(F)$ be the number of $2^{-m}$-mesh cubes that
  intersect $F$.
  It is then shown that
  \begin{equation}
    \overline{\dim}_B F \equiv
    \Limsup_{m \to \infty} \frac{\log_2 M_m(F)}{m}
    \label{eq: bdc}
  \end{equation}
  (see e.g., \cite{Falconer 90}).

  Suppose that $x=(x^1, \dots, x^N)$ is any point in $F$.
  Since the r.e. condition holds for $F$,
  there exists an r.e. set $L$ such that
  $\mathfrak{M}(F) \subset L \subset \hat{\mathfrak{M}}(F)$.
  We consider the following procedure in order to calculate $x_n$.

  Given $n$, one enumerates all elements $(s_1, \dots, s_N)$ of $L$
  such that $\abs{s_1}=\lceil n/N \rceil$.
  There then appears the element $(t_1, \dots, t_N)$
  in the enumeration with the property that
  \begin{equation}
    (x^1\bmod 1, \dots, x^N\bmod 1)\in I(t_1, \dots, t_N).
  \end{equation}
  Assume this $(t_1, \dots, t_N)$ is the $k_n$th element
  in the enumeration order.
  If one knows $n$ and $k_n$,
  then one can calculate the first $\lceil n/N \rceil$ bits of the
  base-two expansion of each $x^i\bmod 1$ with infinitely many zeros
  and hence one can calculate $x_n$ further.

  Thus,
  since $k_n \le 3^N M_{\lceil n/N \rceil}(F)$,
  we see that $H(x_n) \le \log_2 M_{\lceil n/N \rceil}(F) + o(n)$.
  Using \eqref{eq: bdc},
  the result is produced.
\end{proof}

Theorem \ref{ubt} immediately gives the following corollary.

\begin{corollary}[]\label{kolmogorov}
  Suppose that $F$ is a bounded subset of $\R^N$ and satisfies
  the r.e. condition.
  Then $\overline{\dim}_A F \le \overline{\dim}_B F$.
  Moreover,
  for each $k=1,2,3,4$, if $\dim_{Ak}F$ exists
  then $\dim_{Ak}F \le \overline{\dim}_B F$.
\end{corollary}

\begin{theorem}
\label{ead}
  Let $F$ be a bounded subset of $\R^N$.
  Suppose that $F$ satisfies the r.e. condition and
  $\dim_H F=\overline{\dim}_B F$.
  Let $D=\dim_H F$.
  \begin{enumerate}
    \item $\dim_{A4} F$ exists and $\dim_{A4} F = \dim_H F$.
    \item If $\Hm^D(F) > 0$, then $\dim_{A3} F$ exists
      and $\dim_{A3} F = \dim_H F$.
    \item If $\Hm^D(F) > 0$ and $\dim_H F$ is a computable real number,
      then both $\dim_{A1} F$ and $\dim_{A2} F$ exist and
      $\dim_{A1} F = \dim_{A2} F = \dim_H F$.
  \end{enumerate}
\end{theorem}

\begin{proof}
  It follows from Theorem \ref{ubt} that
  $x$ is $(\dim_H F)/N$-compressible for any $x\in F$.
  Using Corollary \ref{ryabko},
  we see that $\dim_{A4} F = \dim_H F$.
  If $\Hm^D(F) > 0$ then, from Corollary \ref{crlbt},
  $\Hm^D(F\backslash T^3_D) = \Hm^D(F) > 0$.
  Hence, $F\backslash T^3_D\neq\phi$,
  therefore there is $x\in F$ which is semi $D/N$-random.
  Thus, we see that $\dim_{A3} F = \dim_H F$.
  Moreover, if $\dim_H F$ is a computable real number,
  then using Theorem \ref{lbt} in a similar manner we see that
  $\dim_{A1} F = \dim_{A2} F = \dim_H F$.
\end{proof}

\begin{theorem}
\label{sas}
  Let $S_1,\dots,S_m$ be contractions on $\R^N$.
  Suppose that each $S_i$ is an affine transformation,
  i.e.,
  for each $i$, $S_i(x)=M_ix+v_i$ where $M_i$ is an $N\times N$
  matrix and $v_i$ is a vector in $\R^N$.
  If all matrix elements of $M_i$ and all components of $v_i$ are
  computable real numbers for each $i$,
  then the r.e. condition holds for the invariant set of $S_1,\dots,S_m$.
\end{theorem}

\begin{proof}
  Let $F$ be the invariant set of $S_1,\dots,S_m$.
  Since $S_1,\dots,S_m$ are contractions on $\R^N$,
  there exists $l\in\N$ such that $S_i(E)\subset E$ for each $i$ where
  $E=
  \left\{\,
    x\in\R^N\bigm|\abs{x}\le l/2
  \,\right\}$.
  We write $S_{i_1, \dots, i_k}=S_{i_1} \circ \dots \circ S_{i_k}$.
  Using Theorem \ref{ifs},
  for each $k$ it is shown that
  \begin{equation}
    F \subset
    \bigcup_{i_1, \dots, i_k} S_{i_1, \dots, i_k}(E)
  \end{equation}
  and $F \cap S_{i_1, \dots, i_k}(E)\neq\phi$ for any $i_1, \dots, i_k$.
  Let $c_1, \dots, c_m$ be the ratios of $S_1, \dots, S_m$ respectively.
  We choose $r \in \Q$ such that $c_i<r<1$ for all $i$ and choose
  $x_0 \in E \cap \Q^N$ such as $(0, 0, \dots, 0)$.
  Then $\abs{S_{i_1, \dots, i_k}(E)} \le lr^k$ for any $k$ and
  $i_1, \dots, i_k$.
  Since all matrix elements of $M_i$ and all components of $v_i$ are
  computable real numbers for all $i$,
  it follows that given $k$, $i_1, \dots, i_k$, and $n\in\N$ one can
  find an $f(n; k; i_1, \dots, i_k)\in\Q^N$ with the property that
  \begin{equation}
    \abs{f(n; k; i_1, \dots, i_k)-S_{i_1, \dots, i_k}(x_0)}
    \le\frac{1}{n}.
  \end{equation}
  It is then shown that
  $\mathfrak{M}(F) \subset L \subset \hat{\mathfrak{M}}(F)$
  holds for the set $L$ accepted by the following procedure,
  and hence $F$ satisfies the r.e. condition.
  
  Given $(s_1, \dots, s_N)\in X^N$,
  one checks whether or not $\abs{s_1} = \dots = \abs{s_N}$ holds true.
  If this does not hold true, then one does not accept
  $(s_1, \dots, s_N)$.
  Otherwise when this does hold true,
  one chooses $k$ such that $lr^k\le\delta/4$
  and chooses $n$ such that $1/n\le\delta/4$,
  where $\delta=2^{-\abs{s_1}}$.
  Let
  \begin{equation}
    \begin{split}
      A&^j(i_1, \dots, i_k) \\
      &=
      [\,
        f^j(n; k; i_1, \dots, i_k)-\delta/2,\,
        f^j(n; k; i_1, \dots, i_k)+\delta/2
      \,]
    \end{split}
  \end{equation}
  where $f^j(n; k; i_1, \dots, i_k)$ is the $j$th component of
  $f(n; k; i_1, \dots, i_k)$,
  and let
  \begin{equation}
    A(i_1, \dots, i_k)=
    A^1(i_1, \dots, i_k) \times \dots \times
    A^N(i_1, \dots, i_k).
  \end{equation}
  It is easy to see that
  \begin{equation}
    F \subset
    \bigcup_{i_1, \dots, i_k} A(i_1, \dots, i_k)
  \end{equation}
  and $F \cap A(i_1, \dots, i_k)\neq\phi$ for any $i_1, \dots, i_k$.
  One then accepts $(s_1, \dots, s_N)$ if and only if
  one can find a $(i_1, \dots, i_k)$ such that
  $I(s_1, \dots, s_N)\cap (A(i_1, \dots, i_k)\bmod 1) \neq \phi$.
\end{proof}

We refer to the following familiar theorem on a self-similar set
(e.g., Theorem 9.3 in \cite{Falconer 90}).
\begin{theorem}\label{sss}
  Suppose that the open set condition \eqref{eq: osc}
  holds for similarities $S_1,\dots,S_m$ on $\R^N$
  with ratios $c_1,\dots,c_m$ respectively.
  If $F$ is the invariant set of $S_1,\dots,S_m$,
  then $\dim_H F = \overline{\dim}_B F = D$ and $0<\Hm^D(F)<\infty$,
  where $D$ is given by
  \begin{equation}
    \sum_{i=1}^m c_i^D = 1. \label{eq: sd}
  \end{equation}
\end{theorem}

\bigskip
The proof of Theorem \ref{main} is as follows.

\begin{proof}[Proof of Theorem \ref{main}]
  Suppose that the open set condition holds for similarities
  $S_1,\dots,S_m$ on $\R^N$ with ratios $c_1,\dots,c_m$ respectively.
  Let $F$ be the invariant set of $S_1,\dots,S_m$.
  From Theorem \ref{sss},
  it is shown that
  $\dim_H F
  = \overline{\dim}_B F
  = D$ and $0<\Hm^D(F)<\infty$,
  where $D$ satisfies
  \eqref{eq: eeq}.
  Furthermore,
  suppose that for each $i$ there exist $N\times N$ matrix $M_i$ and
  $v_i\in\R^N$ such that all matrix elements of $M_i$ and all components
  of $v_i$ are computable real numbers and $S_i(x)=M_ix+v_i$.
  Using Theorem \ref{sas},
  we see that $F$ satisfies the r.e. condition.
  Since each $c_i$ is a computable real number,
  $D$ satisfying \eqref{eq: eeq} is also a computable real number.
  It follows from Theorem \ref{ead} that
  $\dim_{A1} F$ exists and $\dim_{A1} F = \dim_H F$.
\end{proof}

\begin{corollary}
  Let $P=\{p_1,p_2,\dots,p_m\}$ be a finite prefix-free subset of $X$,
  and let $F$ be the set of infinite binary sequences that consist of
  elements of $P$,
  i.e.,
  \begin{equation}
    F =
    \left\{\,
      q_1 q_2 q_3 \dotsc\in X^\infty\bigm|
      q_i\in P\text{ for each }i
    \,\right\}.
  \end{equation}
  Then
  \begin{align}
    &\forall\;\alpha \in F \quad
    \forall\;n \in \N \quad
    H(\alpha_n)\le Dn+o(n),\\
    &\exists\;\alpha \in F \quad
    \lim_{n \to \infty}
    H(\alpha_n)-Dn
    = \infty,
  \end{align}
  where $D$ is given by
  \begin{equation}
    \sum_{i=1}^m 2^{-D\abs{p_i}}=1.
  \end{equation}
\end{corollary}

\begin{proof}
  For each $i$, let $S_i$ be the similarity on $\R$ with
  $S_i(x)=2^{-\abs{p_i}}x+0.p_i$.
  Since $P$ is a prefix-free set, it is shown that $S_1,\dots,S_m$
  satisfy the open set condition.
  Note that all of $2^{-\abs{p_i}}$ and $0.p_i$ are computable real
  numbers.
  Let
  $R(F) =
  \left\{\,
    0.\alpha
    \bigm| \alpha\in F
  \,\right\}$.
  Then $R(F)$ is the invariant set of $S_1,\dots,S_m$.
  From Theorem \ref{main}, we see that $\dim_{A1} R(F) = D$,
  and hence the results are produced.
\end{proof}

\begin{proposition}\label{dim-T}
  Let $D$ be a real number with $0\le D \le N$.
  \begin{enumerate}
    \item $\dim_H T^3_D = D$.
    \item If $D$ is a computable real number
      then $\dim_H T^1_D = \dim_H T^2_D = D$.
  \end{enumerate}
\end{proposition}

\begin{proof}
  Since the results are trivial for $D=0$,
  we assume that $D>0$.

  (a) For any $\varepsilon > 0$,
  we choose a computable real number $d$ such that $D-\varepsilon\le d<D$.
  Noting Theorem \ref{main},
  we can construct similarities $S_1,\dots,S_m$ on $\R^N$ such that
  $\dim_{A3} F = \dim_H F = d$ holds for the invariant set $F$ of
  $S_1,\dots,S_m$.
  Hence, $F\subset T^3_D$ and therefore $D-\varepsilon \le \dim_H T^3_D$.
  Thus, $D\le\dim_H T^3_D$ and using Corollary \ref{crlbt}
  the result is produced.

  (b) From the fact that $T^3_D\subset T^2_D\subset T^1_D$,
  it follows that $D\le\dim_H T^2_D\le\dim_H T^1_D$.
  Since $D$ is a computable real number,
  using Theorem \ref{lbt} we see that $\dim_H T^1_D\le D$.
  Thus, the result is produced.
\end{proof}

Note that Proposition \ref{dim-T} (a) was derived by \cite{Ryabko 84}.

\section*{Acknowledgment}

The author would like to express his special thanks to Prof. Ichiro Tsuda
for a great deal of valuable advice.

\end{document}